\documentclass[11pt]{article}

\usepackage{graphicx,amssymb,amsmath,amsfonts}
\usepackage{hyperref}
\usepackage[margin=1.1in]{geometry}
\usepackage[noadjust]{cite}

\numberwithin{equation}{section}

\usepackage[utf8]{inputenc}

\date{\today}

\def\be{\begin{equation}}
\def\ee{\end{equation}}

\newcommand{\Rea}{\mathrm{Re}}

\newcommand{\avg}[1]{\langle #1 \rangle}

\newmuskip\pFqmuskip

\newcommand*\pFq[6][8]{%
  \begingroup % only local assignments
  \pFqmuskip=#1mu\relax
  \mathchardef\normalcomma=\mathcode`,
  \mathcode`\,=\string"8000
  \begingroup\lccode`\~=`\,
  \lowercase{\endgroup\let~}\pFqcomma
  {}_{#2}F_{#3}{\left(\genfrac..{0pt}{}{#4}{#5}\Big | #6\right)}%
  \endgroup
}
\newcommand{\pFqcomma}{{\normalcomma}\mskip\pFqmuskip}
\newcommand{\pt}[1]{\left(#1\right)}

\newcommand{\pmat}{\begin{pmatrix}}
\newcommand{\fpmat}{\end{pmatrix}}
\newcommand{\eq}{\begin{equation}}
\newcommand{\feq}{\end{equation}}
\newcommand{\cas}{\begin{cases}}
\newcommand{\fcas}{\end{cases}}

\newcommand{\eqarray}{\begin{eqnarray}}
\newcommand{\feqarray}{\end{eqnarray}}
\newcommand{\Tr}[1]{\operatorname{Tr}\pt{#1}}

\newcommand{\diag}[1]{\operatorname{diag}\pt{#1}}

\newcommand{\Imm}{\,\text{Im}\,}

\newcommand{\sff}{\mathrm{SFF}}
\newcommand{\scff}{\mathrm{ScFF}}
\begin{document}
\begin{titlepage}

\title{{{\small{\flushright CCTP-2024-03\\ \hfill ITCP-IPP-2024/3}}} \vspace{2mm}\\ \textbf{From spectral to scattering form factor}\vspace{2mm}}
\author{\textbf{Massimo Bianchi}\(^{[a,b]}\) \\
\href{mailto:massimo.bianchi@roma2.infn.it}{massimo.bianchi@roma2.infn.it} \and
\textbf{Maurizio Firrotta}\(^{[c]}\) \\
\href{mailto:maurizio.firrotta@gmail.com }{mfirrotta@physics.uoc.gr} \and
\textbf{Jacob Sonnenschein}\(^{[d]}\) \\ \href{mailto:cobi@tauex.tau.ac.il}{cobi@tauex.tau.ac.il} \and \textbf{Dorin Weissman}\(^{[e]}\) \\ \href{mailto:dorin.weissman@apctp.org}{dorin.weissman@apctp.org}
}
\date{
\(^{[a]}\)\emph{Dipartimento di Fisica, Universit\`a di Roma Tor Vergata},\\
\emph{Via della Ricerca Scientifica 1, 00133, Roma, Italy} \\[\baselineskip] 
\(^{[b]}\)\emph{INFN sezione di Roma Tor Vergata} \\
\emph{Via della Ricerca Scientifica 1, 00133 Roma, Italy} \\[\baselineskip] 
\(^{[c]}\)\emph{Crete Center for Theoretical Physics, Institute for Theoretical and Computational Physics,
Department of Physics, Voutes University Campus,
GR-70013, Vasilika Vouton, Heraklion} 
\emph{Greece} \\[\baselineskip] 
\(^{[d]}\)\emph{The Raymond and Beverly Sackler School of Physics and Astronomy},\\
\emph{Tel Aviv University, Ramat Aviv 69978, Tel Aviv, Israel}
% \\ [1.3\baselineskip]
% \(^{[d]}\)\emph{Simons Center for Geometry and Physics, SUNY, Stony Brook, NY 11794, USA}
\\ [\baselineskip]
 \(^{[e]}\)\emph{Asia Pacific Center for Theoretical Physics},\\
	\emph{Pohang 37673, Republic of Korea} 
 \\ [\baselineskip]
}
\maketitle
\thispagestyle{empty}

\begin{abstract}
We propose a novel indicator for chaotic quantum scattering processes, the scattering form factor (ScFF). It is based on mapping the locations of peaks in the scattering amplitude to random matrix eigenvalues, and computing the analog of the spectral form factor (SFF). We compute the spectral and scattering form factors of several  non-chaotic systems. We determine the ScFF associated with the phase shifts of the leaky torus, closely related to the distribution of the zeros of Riemann zeta function. We compute the ScFF for the decay amplitude of a highly excited string states into two tachyons. We show that it displays the universal features expected from random matrix theory -  a decline, a ramp and a plateau - and is in general agreement with the Gaussian unitary ensemble. It also shows some new features, owning to the special structure of the string amplitude, including a ``bump'' before the ramp associated with gaps in the average eigenvalue density.
The ``bump" is removed for highly excited string states with an appropriate state dependent unfolding. We also discuss the SFF for the Gaussian $\beta$-ensemble, writing an interpolation between the known results of the Gaussian orthogonal, unitary, and symplectic ensembles.
\end{abstract}

\end{titlepage}

\flushbottom

\tableofcontents
\thispagestyle{empty}

\clearpage
\setcounter{page}{1}
\section{Introduction}
In recent years there has been a resurgence of interest in chaotic systems in many different contexts. In physics they appear both in classical and in quantum
phenomena, both for single-body and for many-body systems. Whereas chaos in classical and quantum mechanical systems has been intensively studied, the chaotic behavior of QFTs and string theories is much less understood.  Chaos in string theory is probably related to Black Holes \cite{Festuccia:2005pi, Festuccia:2006sa} that  have been argued to be fast scramblers \cite{Sekino_2008, Shenker:2013pqa, Balasubramanian_2017} and a bound of chaos has been proposed relying on holography\footnote{Mild violations have been observed near extremality \cite{Bianchi:2020des, Bianchi:2020yzr}.} \cite{Maldacena:2015waa}. 

Chaotic behaviour of scattering amplitudes has been the focus of several investigations in the past couple of years, leading to the realization that highly excited string states may provide a rich setting where to address the issue. Indeed, following the original observations by Gross and Rosenhaus \cite{Gross:2021gsj,Rosenhaus:2020tmv,Rosenhaus:2021xhm}, we have proposed a quantitative measure for the chaotic behaviour of string scattering amplitudes \cite{Bianchi:2022mhs,Bianchi:2023uby}, suggesting the relevance of Random Matrix Theory (RMT) \cite{Mehta:book} and of the so-called $\beta$-ensemble \cite{Baker:1996kq,Forrester:book}.

For these reasons, an important tool for analyzing chaotic systems are random matrices \cite{Mehta:book}. The most familiar in physical systems are the three classical Gaussian ensembles of random Hermitian matrices: the Gaussian orthogonal (GOE), unitary (GUE) and symplectic (GSE) ensembles. The GOE consists of symmetric matrices with entries that are statistically independent  random real variables. The system is invariant under orthogonal transformations. The GUE is an ensemble of hermitian matrices that is invariant under unitary transformations, while the GSE is an ensemble of self-dual hermitian matrices that is invariant under symplectic transformations. 

The Gaussian ensembles are best used to describe systems with random Hamiltonians, with the choice of ensemble depending on the symmetries of the system. Scattering matrices are unitary rather than Hermitian, and as such are better described by the circular ensembles of random unitary matrices. This approach was already taken in the early 1990s in the study of chaotic scattering in quantum mechanics, where the $S$-matrix of some systems was considered as a random unitary matrix \cite{Blumel:1990zz}, but applications to quantum field theory or string theory are scarce.

In analogy with the Gaussian ensembles, one can define the circular orthogonal (COE), unitary (CUE), and symplectic (CSE) ensembles. In many respects they are identical to their corresponding Gaussian ensembles, at least in the limit of large matrices.

Random matrices can be characterized by the statistics of their eigenvalues and eigenvectors. Denoting  the eigenvalues (or eigenphases for a unitary matrix) of a given random matrix by $\lambda_i$ with $i=1,\ldots,L$ where $L$ is the order of the matrix, one can consider the two following important characterizations of the matrix distributions:

(i) The spacings between adjacent eigenvalues  
\be \delta_i = \lambda_i -\lambda_{i+1}\ee
and their ratios 
\be r_i \equiv \frac{ \lambda_{i+1}-\lambda_i}{\lambda_i-\lambda_{i-1}} = \frac{\delta_{i+1}}{\delta_i}
\ee
or the restricted ratios
\be \tilde r_i =\min\{r_i,\frac{1}{r_i}\} \ee
which are defined to be $0\leq \tilde r \leq 1$.

(ii) The random matrix form factor, defined by 
\be\label{RMFF}
\textrm{RMFF}(t)=\frac{1}{L^2}\sum_{i,j} e^{it(\lambda_i-\lambda_j)}
\ee 
where $t\in {\mathbb{R}}$ is a real variable.

\begin{figure}[t!] \centering
    \includegraphics[width=0.48\textwidth]{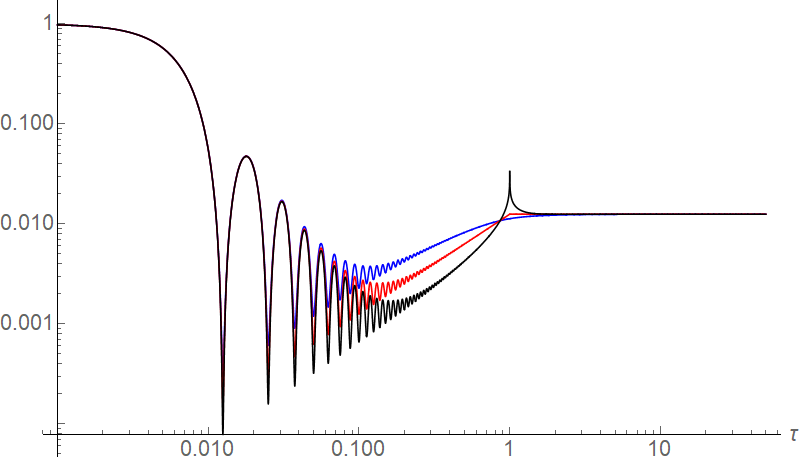} 
    \includegraphics[width=0.48\textwidth]{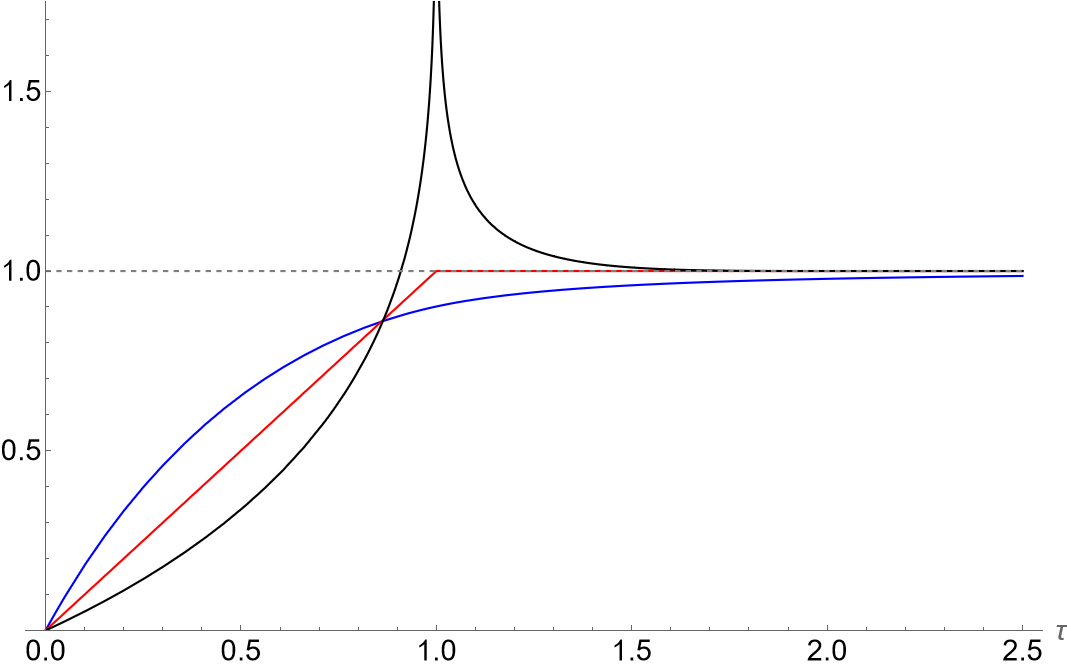} 
    \caption{RMT predictions for the SFF for the classical Gaussian ensembles: GOE (blue), GUE (red), GSE (black). Left is the full SFF on a logarithmic scale (for matrices of size $L=100$, and after unfolding), and right is the connected part of the SFF ($1-r_2(\tau)$) showing the ramp structure on a linear scale. See section \ref{sec:beta} for details.} \label{fig:rmtsff}
\end{figure}

The distribution function $f_\beta(r)$ for the $\beta$
ensemble for $3\times 3$ matrices is given in (\ref{eq:beta_r}). The RMFF appears in figure (\ref{fig:rmtsff}).

In analyzing ``chaotic behavior" of mathematical and physical systems, one first identifies a set of variables that can be mapped to the set $\{\lambda_i\}$ of the random matrix eigenvalues, in the sense that the analogs of the ratios (i) and of the form factor (ii) admit the same universal properties as those of the random matrices. If the symmetries of the system are known, one can anticipate which of the RMT ensembles would be appropriate for the system at hand.
One such very well known map, due to Wigner and Dyson \cite{Dyson:1962b} is to  the set of the eigenvalues of the Hamiltonian of a given chaotic system, namely the spectrum of that system. In this case the map is 
\be
E_i:\quad   {i=1,... L}\qquad  \rightarrow \qquad  \lambda_i = I(E_i) :\quad  {i=1,...L}
\ee
where $L$ for the spectral case is the dimension of the Hilbert space. The function $I(E)$ used in this map is determined from the average eigenvalue density function $\rho(E)$. One should first \emph{unfold} the spectrum in such a way that the average density in the new variable is constant, and the behavior of the ``fluctuations'' around the average density $\rho(E)$ is exposed. Correspondingly one defines 
\begin{equation} \delta_i = I(E_{i+1}) - I(E_{i}) \end{equation}
The ratios of successive spacings typically depend only weakly on the unfolding, and therefore one can use directly the spacing ratios defined by
\begin{equation} \label{eq:rn} r_i \equiv \frac{E_{i+1}-E_i}{E_i-E_{i-1}} = \frac{\delta_{i+1}}{\delta_i} \end{equation}
The \emph{spectral form factor } (SFF) can be also defined by 
\be
\sff(t)=\frac{1}{L^2}\sum_{i,j} e^{it(E_i-E_j)} \label{eq:spectralff}
\ee 
For this case the variable $t$ is naturally identified with physical time, and in fact one can relate the SFF to the square modulus of the trace of the evolution operator as described in section \ref{sec:generalproperties}. By taking $t\to t+i\tilde \beta$ one can consider the system at finite temperature\footnote{We use the symbol $\tilde\beta$ for the inverese temperature since later on $\beta$ will be used to denote the parameter of the $\beta$ ensemble.} $T=1/\tilde\beta$ and get a (generalized) partition function. The SFF can typically be separated into a ``disconnected'' part that depends only on the average density $\rho(E)$, and a ``connected'' part, which depends on the fluctuations and will have the distinctive ramp structure for chaotic systems.

Another well known map to the set of eigenvalues of random matrices, is the map from the set of zeros of certain mathematical functions.  If we denote by $z_i$ the zero of the chosen function then the map reads
\be
z_i\   {i=1,... n_z}\qquad  \rightarrow \qquad  \lambda_i\  {i=1,...n}
\ee
where now $n_z$ is the number of zeros. The most famous map of this form is the map of the (non-trivial) zeros of the Riemann zeta function. 

In \cite{Bianchi:2022mhs} and \cite{Bianchi:2023uby} a novel map was proposed between 
 the locations of the extrema of a scattering amplitude denoted  $z_i$ ${i=1,... n}$ where now $n$ stands for the number of extrema  of the amplitude.
\be
{\cal A}(\alpha) \qquad {\cal F}(\alpha)\equiv \frac{d log{\cal A}}{d \alpha}
\ee
namely  the zeros of its logarithmic derivative.
\be
{\cal F}(z_{i})=0
\ee
In \cite{Bianchi:2022mhs} and \cite{Bianchi:2023uby} we supplemented this map with a novel measure of a chaotic behavior of a scattering process in the form of the ratio of spacings
\be \delta_n = z_n - z_{n+1}\ee
and
\be r_n \equiv \frac{z_{n+1}-z_n}{z_n-z_{n-1}} = \frac{\delta_{n+1}}{\delta_n}\,,\qquad \tilde r_n =min\{r_n,\frac{1}{r_n}\} \ee

Our present goal is to investigate the chaotic behavior of scattering processes using the analog of the random matrix form factor that we christen the {\it Scattering form factor}(ScFF).
\be
\mathrm{ScFF}(s)=\frac{1}{n^2}\sum_{z_i,z_j} e^{is(z_i-z_j)} \label{eq:scatteringff}
\ee 
by consonance with the {\it Spectral form factor}.

From the very definition it follows that $\scff(s=0)=1$ and for $s,z_i\in{\mathbb{R}}$ also $\scff(s)\in{\mathbb{R}}$.

Of course, mathematically there is no difference between the expressions in eqs. \eqref{eq:spectralff} and \eqref{eq:scatteringff}. There is only a difference in the physical interpretation and meaning of this function. One key aspect is that the eigenvalues in the ScFF are not the energies. We will consider the ScFF for the angle ``eigenvalues" in highly excited string amplitudes, where $z_i$ are the positions of peaks in the angular dependence of the amplitude. Then, the variable $s$ cannot be identified with time. The physical meaning of $s$ relates to angular momentum as we will argue later on.

\begin{figure} \centering
    \includegraphics[width=0.88\textwidth]{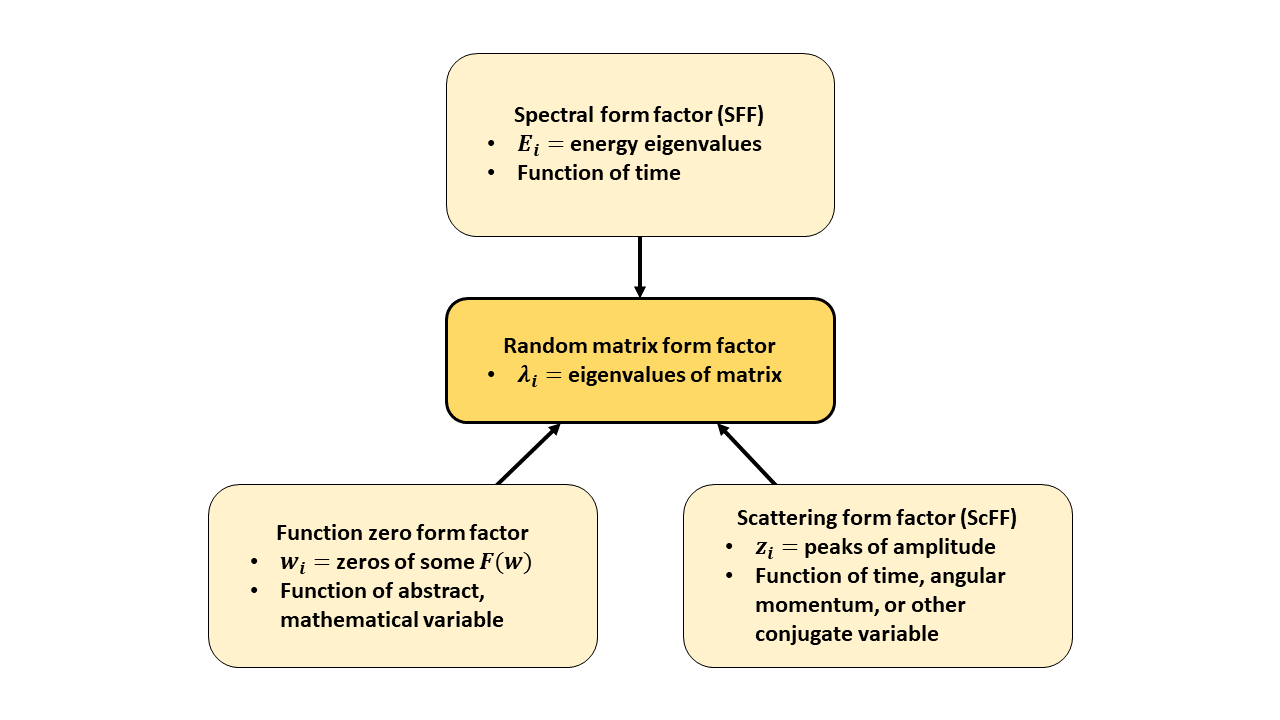}
    \caption{Various different form factors can be mapped to the random matrix form factor, with different quantities related to the matrix eigenvalues. The physical interpretation of the variable that the form factor depends on varies correspondingly. The mapping typically requires some ``unfolding'' to uncover universal properties of RMT. \label{VariousSFF}}
\end{figure}

In figure (\ref{VariousSFF}) we summarize the various maps of the (RMFF) to the spectral form factor SFF, to the function zeros form factor FZFF and to the scattering form factor ScFF.
%%%%%%%%%%%%%%%%%%%%%%%%%%%%%%

%where $E$ and $E'$ are the energy eigenstates in  a window that contains $n$ states centered around the temperature of the system.

%In RMT one defines a similar expression
%\be
%SFF(t)=\frac{1}{n^2}\sum_{i,j'}^n e^{it(\lambda_i-\lambda_j)}
%\ee 
%where $\lambda_i$ are the eigenvalue of the random matrix scaled in such a way that they span the interval $[-1, 1]$.

The SFF is expected to display some `universal' behaviour for chaotic systems \cite{Cotler:2016fpe}.  The so-called D-R-P (Dip-Ramp-Plateau) consists of an initial `decline' for `small' $t$ until a minimum (`dip') is reached a $t=t_d$, a subsequent raise (`ramp') and finally an almost constant behaviour (`plateau') for large $t$. Notwithstanding some existing `non-chaotic' but to some extent artificial counter-examples \cite{Das:2023yfj, Das:2023ulz, Das:2022evy} that display some of the above features, D-R-P is considered a hallmark of chaotic systems. 

%Following the analogy between the (chaotic) distributions of the (normalized) ratioes of (energy) eigenvalues and the (chaotic) distributions of the (normalized) ratioes of extremum points  of a scattering amplitude \cite{StringChaos1, StringChaos2} we propose the analogous ScFF (scattering form factor)
%\be
%ScFF(s)=\frac{1}{n^2}\sum_{z_i,z_j} e^{is(z_i-z_j)}
%\ee 
%where $z_i$ and $z_j$ are the locations of the extrema of a scattering amplitude,  
%\be
%{\cal A}(\alpha) \qquad {\cal F}(\alpha)\equiv \frac{d log{\cal A}}{d \alpha}
%\ee
%namely  the zeros of its logarithmic derivative.
%\be
%{\cal F}(z_{ii})=0
%\ee
%Here $n$ is the number of peaks centered around the average value of the peaks.

Once introduced the ScFF and  its general properties, we   address  in this paper the following topics: (i) The computation of the spectral and scattering form factors of certain non-chaotic systems. These include the SFF of the harmonic oscillator and free open string and the ScFF of the decay of the leading Regge states into two tachyons.
(ii) We determine the ScFF of the leaky torus and relate it to the non-trivial zeros of Riemann z functions.  (iii) The ScFF associated with the decays of HESS into two tachyons including the identification of the decline, ramp and plateau regions in the $s$ variable.

Chaotic scattering in string theory has been investigated also in 
\cite{Dodelson:2022eiz,Firrotta:2022cku, Firrotta:2023wem, Dodelson:2023vrw, Bianchi:2011se, Savic:2024ock, Das:2023cdn, Djukic:2023dgk, Dutta:2023yhx, Bianchi:2016bgx,Hashimoto:2022bll}

The plan of the paper is as follows.

In section 2 we describe  the basic properties of the ScFF, in analogy with the ones of the RMFF and  SFF. This includes  the three regions  of the form factor, namely, the decline, the ramp and the plateau.  

In section 3, we discuss the $\beta$ ensemble of random matrices and  together with its scattering analog.
In section 4 we discuss the spectral and scattering form factors of non-chaotic systems. These include the harmonic oscillator, free string and single massive spin  decay into two scalars, in order to get some insights on the interpretation of the auxiliary variable $s$.  

In section 5, we compute the ScFF for the leaky torus and show its relations with the FZFF of the Riemann zeta function. 

Section 6 is devoted to the ScFF of string 3-point scattering processes. We write down the density of the  maximum points of the scattering amplitude and perform on it a unfolding procedure that allows a successful fit with CUE predictions, including the removal a peculiar `bump' structure that appears before unfolding. 

The summary of the results and several open questions are presented in section 7.

In  appendix A  we describe the integer partitions and structure of peaks in the HES decay amplitude. Appendix B is devoted to the low level spin decomposition and in appendix C we propose a model to account for the ``bump" in the ScFF.

%computation of the 3-point amplitude and present the corresponding spin decomposition.

%we present elementary (integrable) examples whereby SFF/ScFF can be computed (semi) explicitly (ie harmonic oscillator,  decay , free strings, ...) or special cases that can be addressed with well-developed tools (Riemann zeroes, leaky torus, ...). We then pass to consider   
%%the ScFF for the 2-body decay of a highly-excited DDF state and finally (?)
 %for the  scattering (3-body decay).
 
%\clearpage

\section{Spectral form factor in random matrix theory: Predictions for the \texorpdfstring{$\beta$}{beta}-ensembles} \label{sec:beta}
The aim of this section is to summarize the predictions for the spectral form factor coming from random matrix theory (RMT). We begin with the standard classical Gaussian ensembles, and their generalization to the Gaussian $\beta$-ensemble, where $\beta$ is a continuous parameter interpolating between the three classical ensembles, located at $\beta = 1$ (GOE), 2 (GUE), or 4 (GSE).

Our interest in the $\beta$-ensemble is due to the result of \cite{Bianchi:2023uby}, where it was found that the spacing ratios of peaks in the scattering amplitudes of highly excited strings was best fitted with values of $\beta$ between 1 and 2. We offer here another method to measure this parameter by analyzing the spectral form factor, and comparing with RMT predictions. We present (numerical) results for the SFF for continuous values of $1\leq \beta \leq 4$, with a proposed formula for the interpolation between the GOE, GUE and GSE results. These results will be also valid for the circular ensembles of random unitary matrices, as is also discussed in the following, and therefore could be applicable to a large class of physical models.

Though we will focus on the region $1\leq\beta\leq4$ in this section, we should remark that there is also a formal connection of the $\beta$ ensemble with the non-chaotic Poisson distribution. Properly taking the limit $\beta\to0$, one finds Poisson statistics at the edge of Gaussian $\beta$ ensemble at high temperature \cite{2018arXiv180408214P}.

\subsection{The Gaussian \texorpdfstring{$\beta$}{beta}-ensemble}
The Gaussian $\beta$-ensemble (GBE) can be defined starting from the joint probability distribution function of the eigenvalues,
\be P_L(\lambda_1,\lambda_2,\ldots,\lambda_L) = {\cal N}_L(\beta) \times \exp\left(-\frac\beta2\sum_{i=1}^L \lambda_i^2\right) \prod_{1\leq i < j \leq L} |\lambda_i-\lambda_j|^\beta \label{eq:beta_lambda} \ee
with the normalization constant given by \cite{Forrester:book}
\be {\cal N}_L(\beta) = \frac{\beta^{\frac12L + \frac14\beta L(L-1)}}{(2\pi)^L)}\prod_{j=1}^{L} \frac{\Gamma(1+\frac12\beta)}{\Gamma(1+\frac12 j\beta)} \ee

The three classical Gaussian ensembles, orthogonal (GOE), unitary (GUE) and symplectic (GSE), correspond to the values $\beta = 1$, 2, and 4, respectively. However, one can study the properties of the $\beta$ ensemble starting from the distribution \eqref{eq:beta_lambda} for any real $\beta>0$. Physically, it is interpreted as the statistics of a one dimensional Coulomb gas of $L$ charged particles with a logarithmic interaction, and $\beta$ is exactly the inverse temperature of the gas \cite{Dyson:1962es,Dyson:1962b,Forrester:book}.

A convenient and widely used matrix model for the GBE is the tridiagonal construction of Dumitriu and Edelman \cite{Dumitriu:2002}, which defines the ensemble in terms of symmetric matrices with real random entries. The $L\times L$ matrices have the tridiagonal form
\be {\cal M}_\beta  \sim \frac{1}{\sqrt{\beta}}\left(\begin{matrix}
N(0,2) & \chi_{(L-1)\beta} & & & \\
\chi_{(L-1)\beta} & N(0,2) & \chi_{(L-2)\beta} & &  \\
& \ddots & \ddots & \ddots & \\
& & \chi_{2\beta} & N(0,2) & \chi_\beta \\
& & & \chi_\beta & N(0,2)
\end{matrix}\right) \ee
This notation is rather schematic. The meaning is that the diagonal elements are $L$ independent Gaussian variables drawn from a normal distribution with $\mu=0$ and $\sigma^2=2$. The elements on the next diagonals are drawn from a $\chi$-distribution with parameter $k\beta$, with $k = L-1,L-2,\ldots,1$, going down the diagonal. Since the matrix is taken to be symmetric, there are $L-1$ independent variables of this kind.

In \cite{Dumitriu:2002} it was proven that the eigenvalues of these matrices obey the statistics of \ref{eq:beta_lambda}. It is particularly useful since it is easy and quick, from a computational point of view, to obtain GBE spectra by generating random matrices of the above form and diagonalizing them.

The spectra thus obtained will also obey the semicircle law. That is, the average density of eigenvalues is 
\be \rho(\lambda) = \frac{1}{2\pi L}\sqrt{4L-\lambda^2} \label{eq:semicirclerho}\ee
where $L$ is the size of the matrix (which we usually assume is large). After defining the cumulative distribution as
\be I(\lambda) = L\int_{-2\sqrt{L}}^\lambda d\lambda^\prime \rho(\lambda^\prime) = 
    % \begin{cases} 0 & E \langle -2\sqrt{N} \\
            \frac{L}{2} + \frac{L}{\pi}\arctan\left(\frac{\lambda}{\sqrt{4L-\lambda^2}}\right) + \frac{\lambda\sqrt{4L-\lambda^2}}{4\pi} 
            %& -2\sqrt{N}\leq E \leq 2\sqrt N 
            % \\            1 & E \rangle 2\sqrt N \end{cases}
            \label{eq:semicircleI}
\ee
then for the \emph{unfolded spectrum} defined as \be z_n \equiv I(\lambda_n) \ee
the average density is constant between $0$ and $L$. The normalized level spacings $\delta_n \equiv z_{n+1}-z_n$ have a mean of $\avg{\delta_n}=1$, and obey the usual Wigner-Dyson distributions,
\be p_{\beta}(\delta) = {\cal C}_{\beta} \delta^\beta \exp(-c_\beta \delta^2) \label{eq:beta_delta} \ee
with
\be {\cal C}_\beta = 2\frac{[\Gamma(\frac{\beta+2}2)]^{\beta+1}}{[\Gamma(\frac{\beta+1}2)]^{\beta+2}}\,,\qquad c_\beta = \left(\frac{\Gamma(\frac{\beta+2}2)}{\Gamma(\frac{\beta+1}2)}\right)^2\ee

The spacing ratios $r_n \equiv \frac{\delta_{n+1}}{\delta_n}$ for $3\times 3$ random matrices are distributed according to \cite{Atas:2013dis}
\be f_\beta(r) = \frac{3^{\frac{3+3\beta}2}\Gamma(1+\frac\beta2)^2}{2\pi \Gamma(1+\beta)} \frac{(r+r^2)^\beta}{(1+r+r^2)^{1+\frac32\beta}} \label{eq:beta_r}\ee
In general, if $I(\lambda)$ is a slowly varying function on the scale of $\avg{\delta \lambda}$, then the distribution of $r$ would be nearly the same whether one unfolded the spectrum or not. The spacings of peaks in the amplitude of highly excited strings will provide a counter example of this, as we show in section \ref{sec:unfolding}, because of the presence of `repulsive' points.

\subsection{The circular \texorpdfstring{$\beta$}{beta}-ensemble}
A chaotic scattering matrix is expected to be a random unitary matrix drawn from the Gaussian (unitary) ensemble. The eigenvalues of a unitary matrix all lie on the unit circle, $\lambda_n = e^{i\theta_n}$. 

Random unitary matrices are described by the `circular' ensembles. The distribution of the \emph{eigenphases} $\theta_i$ is given by
\be P(\theta_1,\theta_2,\ldots \theta_L) = \widetilde{\cal N}_L(\beta) \prod_{1\leq i < j \leq L}|e^{i\theta_i}-e^{i\theta_j}|^\beta \label{eq:lambda_cbe} \ee
with
\be \widetilde{\cal N}_L(\beta) = \frac{1}{(2\pi)^L} \frac{[\Gamma(1+\frac12\beta)]^L}{\Gamma(1+\frac12\beta L)} \ee

Like for the Gaussian ensembles, `circular' ensembles describe the statistical behavior of the Coulomb gas at inverse temperature $\beta$, the difference being that now all the charges are placed on the unit circle. The eigenphases are (on average) uniformly distributed on the circle - as one would expect given the rotation symmetry - with a constant density of $\frac{1}{2\pi}$.

It is well known \cite{Mehta:book} that for large matrices, the distribution of spacings of eigenphases in the circular ensembles,
\be \delta_n \equiv \frac{L}{2\pi}(\theta_{n+1}-\theta_n) \ee
goes to the same Wigner-Dyson distribution as for the corresponding Gaussian ensembles.

As with the Gaussian ensembles, the most familiar results are obtained for the special values of $\beta = 1$ (circular orthogonal ensemble/COE), $\beta = 2$ (unitary/CUE), and $\beta = 4$ (symplectic/CSE). Similarly to the Gaussian $\beta$-ensemble, we can define a consistent circular $\beta$-ensemble (CBE) for any $\beta> 0$ starting from \eqref{eq:lambda_cbe}.

For generic values of $\beta$ there is a matrix model for the CBE due to Killip and Nenciu \cite{Killip:2004}. The construction is slightly more involved than in the Gaussian case and relies on the use of Cantero-Moral-Velazquez (CMV) matrices \cite{Cantero:2003}.

Given $\beta > 0$, in order to define a random $L\times L$ unitary matrix from the CBE, we first define $L$ independent complex random variables 
\be \alpha_k \equiv \sqrt{\rho_k} e^{i\phi_k} \ee
for $k=0,1,\ldots,L-1$, such that
the phases $\phi_k$ are drawn from a uniform distribution,
\be \phi_k \sim U(0,2\pi) \ee
and the $\rho_k$ are drawn from the Beta-distribution\footnote{We use a standard definition in which for variable $X\sim B(s,t)$, the PDF is $\propto x^{s-1} (1-x)^{t-1}$ for $0< x < 1$. Note this and other small differences in notation from \cite{Killip:2004}.} as
\be \rho_k \sim B\left(1,\frac12(L-k-1)\beta\right) \ee
with the exception of $k=L-1$, for which one takes $\rho_{L-1} \equiv 1$.
Then, one defines the matrices
\be \Xi_{k} \equiv \left(\begin{matrix}
    \alpha_k^* & \sqrt{1-|\alpha_k|^2} \\
    \sqrt{1-|\alpha_k|^2} & -\alpha_k
\end{matrix}\right)\ee
for $k=0,1,\ldots,L-2$, and
\be \Xi_{-1} \equiv \big(-1\big)\,,\qquad \Xi_{L-1} \equiv \big(\alpha_{L-1}^*\big) \ee
Using these, one defines two block diagonal matrices
\be M_1 \equiv \diag{\Xi_{-1},\Xi_{1},\Xi_3,\ldots}\,, \qquad M_2 \equiv \diag{\Xi_0,\Xi_2,\Xi_4,\ldots} \ee
Finally, the product $M_1 M_2$ (as well as $M_2 M_1$) is a random $L\times L$ unitary matrix whose eigenvalues follow exactly the distribution \eqref{eq:lambda_cbe}, for generic values of $\beta>0$.

Using this definition for the generic circular $\beta$-ensemble, we can verify that numerical results for the spacing and spacing ratio distributions - as well as for the SFF - are indistinguishable between the circular and Gaussian ensembles with the same value of $\beta$, not only for the classical values of 1, 2, 4, but for any values of $\beta$ in between. Therefore all the results cited below in this section will be relevant for both computations involving the eigenvalues of random Hermitian matrices, as well as the eigenphases of random unitary scattering matrices.

\subsection{The spectral form factor for the \texorpdfstring{$\beta$}{beta}-ensembles}\label{sec:rmtsff}
In the same manner that the distributions of spacings and spacing ratios interpolate between the classical GOE, GUE, and GSE when $\beta$ is varied continuously between 1, 2, and 4, we find that the spectral form factor for the GBE offers a smooth interpolation between the SFFs of the three ensembles. To reiterate the point made in the last subsection, we refer below only to the Gaussian ensembles purely for the sake of brevity, since the results are equally applicable to the circular ensembles with the same values of $\beta$.

Let us repeat here some known RMT results.\footnote{Here we follow closely \cite{Liu:2018hlr} for its clear presentation, though the results are well known and found in many other references}. We use the definition
\be \sff(t) = \big\langle\,\frac{1}{L^2}\sum_{i=1}^L\sum_{j=1}^L e^{i(z_i-z_j)t}\,\big\rangle \ee
The averaging is done with respect to the probability function \eqref{eq:beta_lambda}, and so will correspond to an ``ensemble average'' of the SFF.

To be consistent throughout this paper, and for universal applicability, we assume an unfolded spectrum for which the eigenvalue density is constant $\rho(z) = \frac{1}{L}$ between 0 and $L$. With this normalization the SFF starts from 1 at $t=0$, and it will approach the plateau, $\sff \to \frac{1}{L}$, when $t \approx 2\pi$. We also define
\be \tau \equiv \frac{t}{2\pi} \ee
to bring the expressions below to their simplest form, where all constant factors have been absorbed into definitions.

\begin{figure} \centering
   \includegraphics[width=0.48\textwidth]{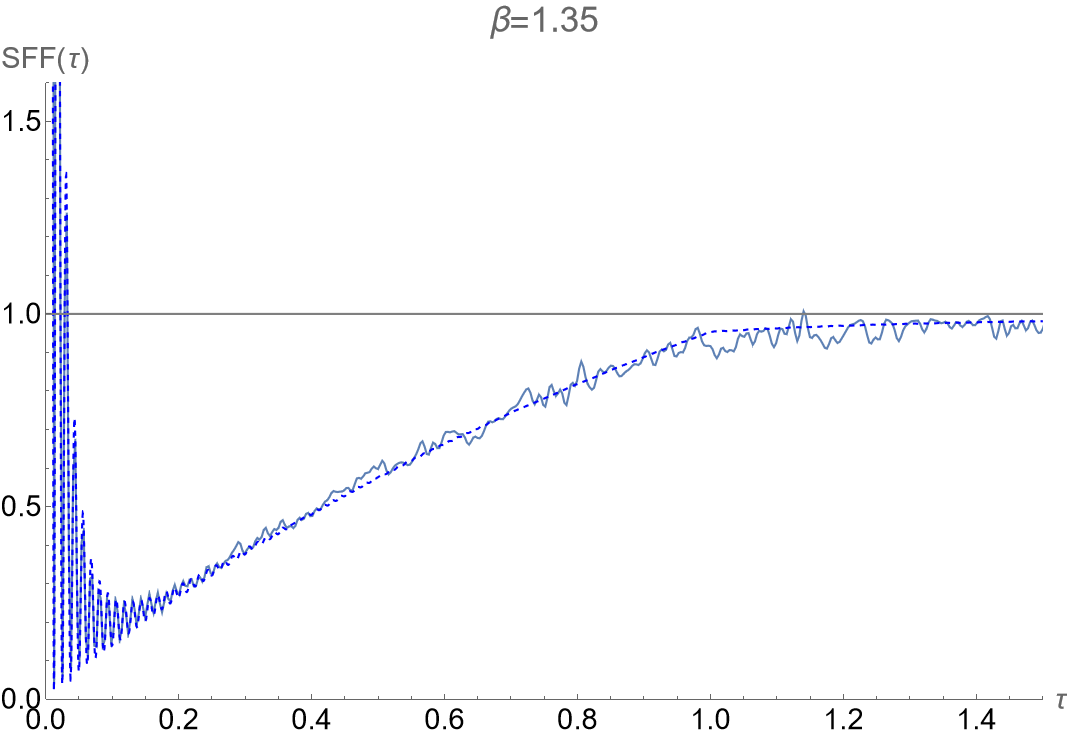}
   \includegraphics[width=0.48\textwidth]{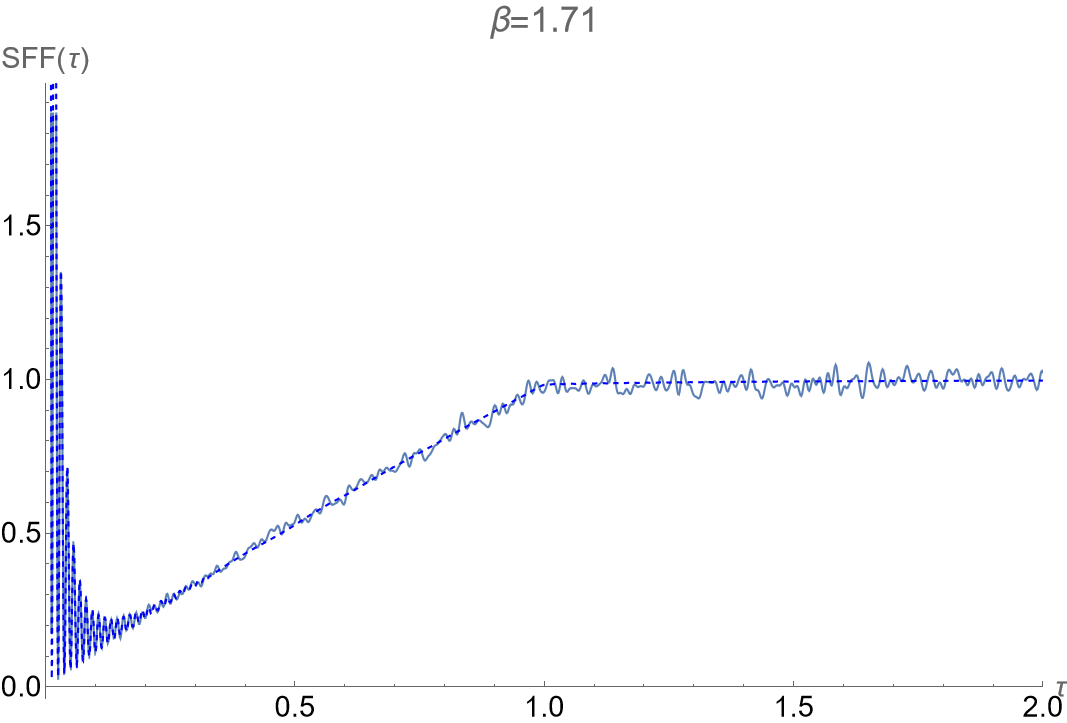} \\
   \includegraphics[width=0.48\textwidth]{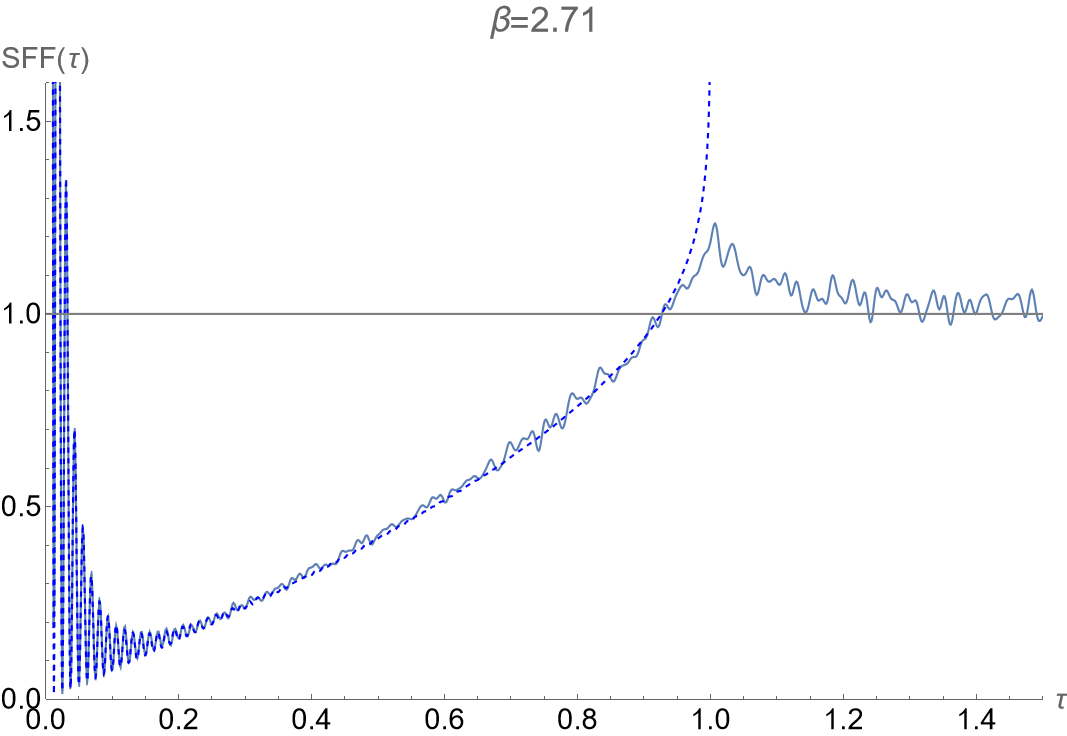}
   \includegraphics[width=0.48\textwidth]{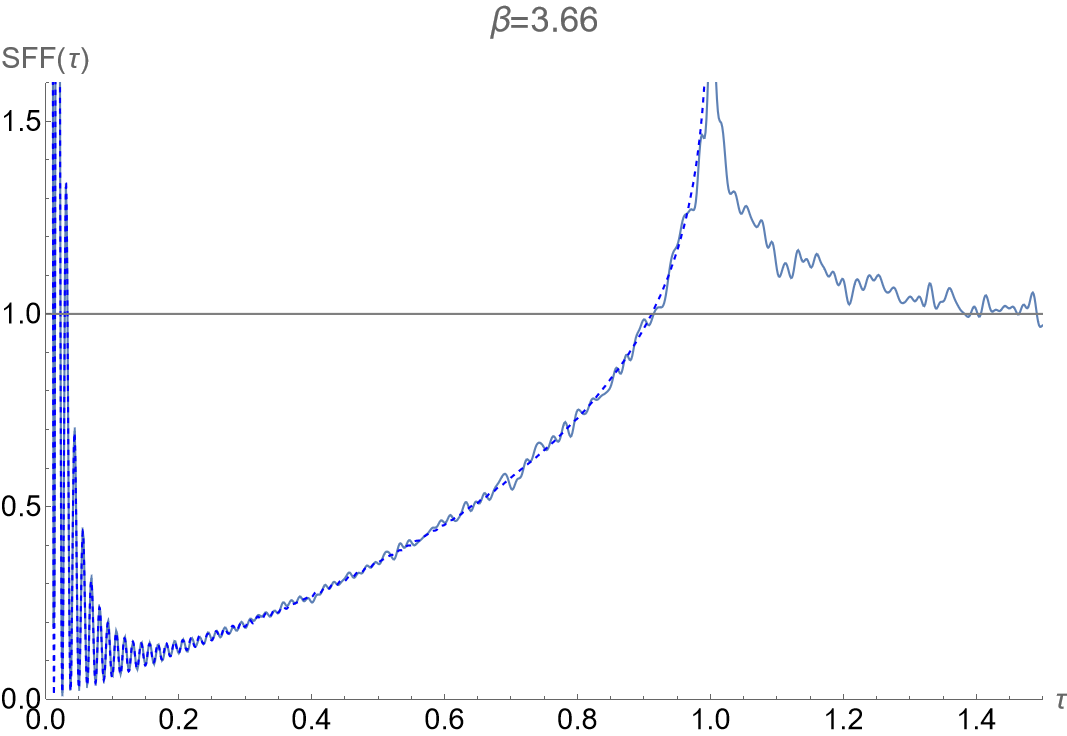}
   \caption{SFF for $\beta$ between 1 and 2 (top), and 2 and 4 (bottom), with our predicted interpolations \eqref{eq:sff_beta}-\eqref{eq:sff_beta2} The result is an ensemble average over 2000 matrices of size $L=80$. \label{fig:sff_beta}}
\end{figure}

In all cases the SFF can be written as
\be \sff(\tau) = r_1^2(\tau) + \frac{1}{L}(1 - r_2(\tau)) \label{eq:sff_rmt} \ee
where $r_1$ is the disconnected part, independent of $\beta$. It is simply the Fourier transform of the eigenvalue density function, which for a constant density is given by
% \be r_1(t) \equiv \frac{J_1(2t)}{t} \ee
\be r_1(\tau) \equiv \frac{\sin(\pi L \tau)}{\pi L \tau} \ee
The ``connected'' part of the SFF determines the ramp and is given by
\begin{align} \label{eq:sff_goe} r_2^{\mathrm{GOE}}(\tau) &= 
    \begin{cases} 1 - 2\tau + \tau\log\left(1+2\tau\right) \qquad & \tau \leq 1 \\
    -1 + \tau\log\left(\frac{2\tau+1}{2\tau-1}\right) & \tau > 1 \end{cases} \\ \label{eq:sff_gue}
    r_2^{\mathrm{GUE}}(t) &= 
    \begin{cases} 1 - \tau  \qquad & \tau \leq 1 \\
    0 & \tau > 1 \end{cases} \\
   \label{eq:sff_gse} r_2^{\mathrm{GSE}}(t) &= 
    \begin{cases} 1 - \frac{\tau}{2} + \frac{\tau}{4}\log|1-\tau| \qquad & \tau \leq 2 \\
   0 & \tau > 2 \end{cases}
\end{align}
The slope of the linear term is seen to be proportional to $\frac{1}{\beta}$. These functions are plotted in figure \ref{fig:sff_beta}.

Although we do not prove it analytically, we find that the following formulae are consistent with numerical results for GBE spectra with $1\leq \beta \leq 4$, and offer a smooth interpolation between the formulae cited. We have to distinguish two cases. The interpolation between GOE and GUE, valid for $1\leq\beta\leq2$ is
\be r_2^{\mathrm{OU}}(\tau) = \begin{cases} 1-\frac{2}{\beta}\tau + \left(\frac{2}{\beta}-1\right)\tau \log(1+2\tau) & \tau < 1 \\
1-\frac{2}{\beta} + \left(\frac{2}{\beta}-1\right)\tau \log\left(\frac{2\tau+1}{2\tau-1}\right) & \tau \geq 1 \end{cases} \label{eq:sff_beta}
\ee
And is found to be in excellent agreement with numerics.

The interpolation from GUE to GSE ($2\leq\beta\leq4$) is somewhat complicated by the fact that the GSE SFF has a log singularity at $\tau = 1$, which gradually develops when $\beta$ is below 4. We find that in the region $\tau<1$, which is the ramp of the SFF, the formula
\be
 r_2^{\mathrm{US}}(\tau) = 1-\frac{2}{\beta}\tau +\frac12\left(1-\frac{2}{\beta}\right) \tau \log|1-\tau| \label{eq:sff_beta2}
\ee
matches well the numerical results, except when very close to $\tau=1$. See figure \ref{fig:sff_beta}.

We can see from the above the linear term in the ramp of the SFF with the slope
\be \mathrm{ramp}(t) = \frac{2\tau}{\beta} = \frac{t}{\pi \beta} \ee
is common to all values of $\beta$ (at least in the interpolating region between 1 and 4). However, the purely linear behavior holds true for GUE only. While for values of $\tau \ll 1$ the linear term dominates (at least after subtracting the disconnected part of the SFF), non-linear terms are significant at later times. For $1\leq\beta < 2$ the connected SFF is a concave function, and it is convex for $2<\beta\leq4$. This can be a useful indicator whether the SFF one has computed corresponds to a value of $\beta$ below or above 2.

One can also repeat these calculations in the region $0<\beta<1$, which interpolates between Poisson statistics and the GOE. The behavior of $2\tau/\beta$ at small $\tau$ can be seen for values of $\beta<1$ as well. For small $\beta < 0.15$ the numerics begin to show some deviations, already at the level of the distribution of spacings. For $\beta=0.3$ we can still verify that the slope of the ramp is exactly $2/\beta$ near the origin.

\subsubsection{SFF for mixed ensembles}
Finally, we discuss one more scenario that could lead to fractional values of $\beta$ in a physical system, which is a mixed ensemble. Let us assume that we have a set of random matrices such that a fraction $p$ is drawn from the GOE, and the remaing fraction $1-p$ is from the GUE. For simplicity, we do not include GSE in this discussion.

Since the interpolation of \eqref{eq:sff_beta} turns out to be linear, the SFF of the mixed ensemble would be indistinguishable from that of a $\beta$-ensemble, since, after ensemble averaging we will get that the SFF is just the sum (writing only the ramp section)
\be \sff_p(\tau) = p\sff_{\mathrm{GOE}} + (1-p)\sff_{\mathrm{GUE}} = (1+p)\tau - p \tau \log(1+2\tau) \ee 
which means that the SFF will look the same as for an effective $\beta$ given by
\be 1 + p = \frac{2}{\beta_p}\,,\qquad \beta_p = \frac{2}{1+p} \ee

On the other hand, if we compute the distribution of the spacing ratios $r_n$ in such a case, we will also get a value of $\beta$ between 1 and 2, but not the same one. One can show that the average $\avg{\tilde r_n}$ in this case,
\be \avg{\tilde r}_p = p \avg{\tilde r}_{GOE} + (1-p)\avg{\tilde r}_{GUE} \ee
will be different from the expected value of $\avg{\tilde r}$ in the $\beta$-ensemble with $\beta=\beta_p = \frac{2}{1+p}$. The differences however are not large.

If one can measure $\beta$ accurately enough from both the distribution of spacing ratios and the SFF, one can distinguish between the scenario where the ensemble is a pure $\beta$-ensemble, in which the value of $\beta$ should be the same from both measurements, and the option of a mixed ensemble, where they would disagree.

\begin{figure} \centering
   \includegraphics[width=0.48\textwidth]{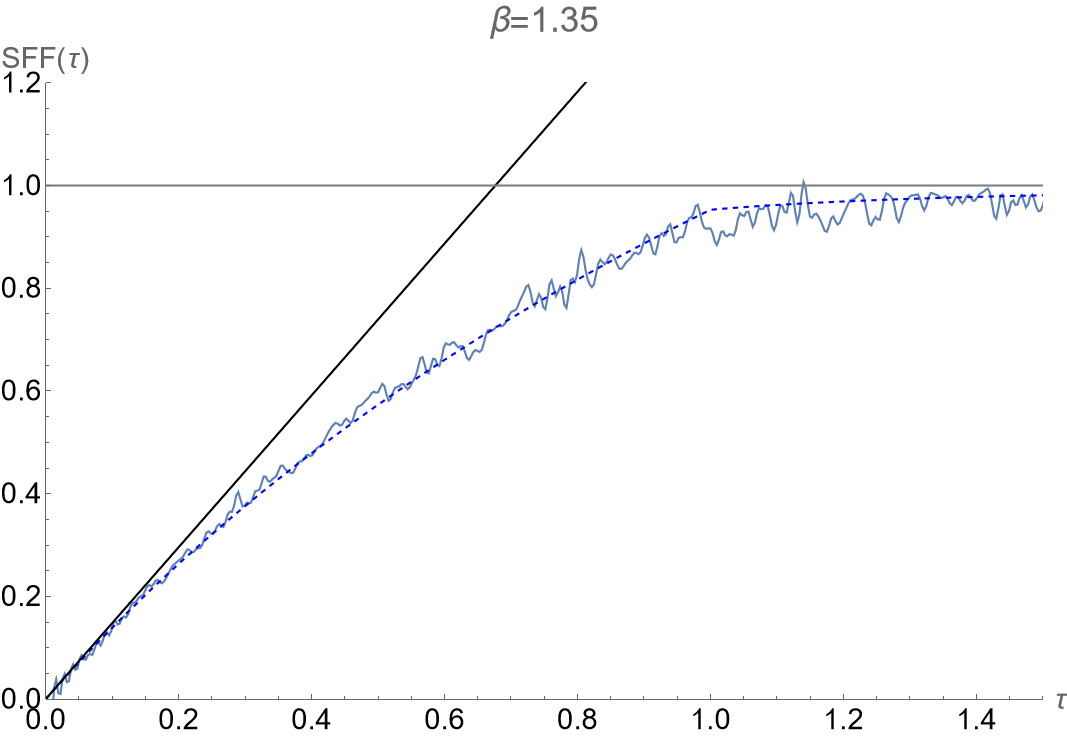}
   \includegraphics[width=0.48\textwidth]{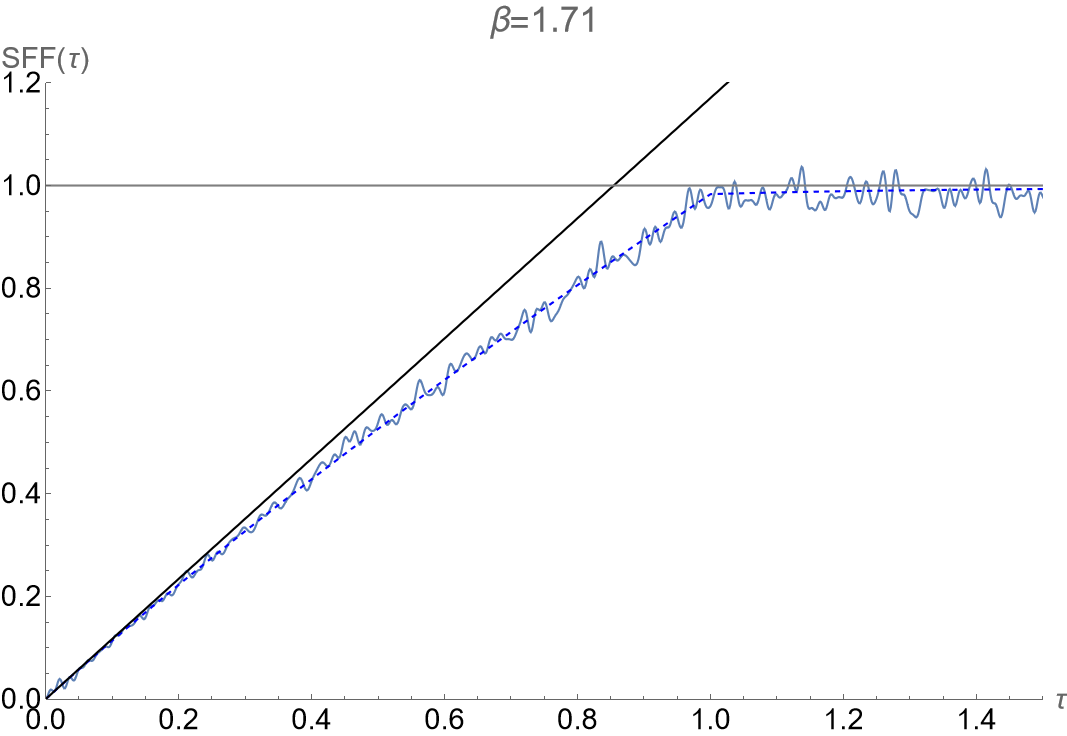} \\
   \includegraphics[width=0.48\textwidth]{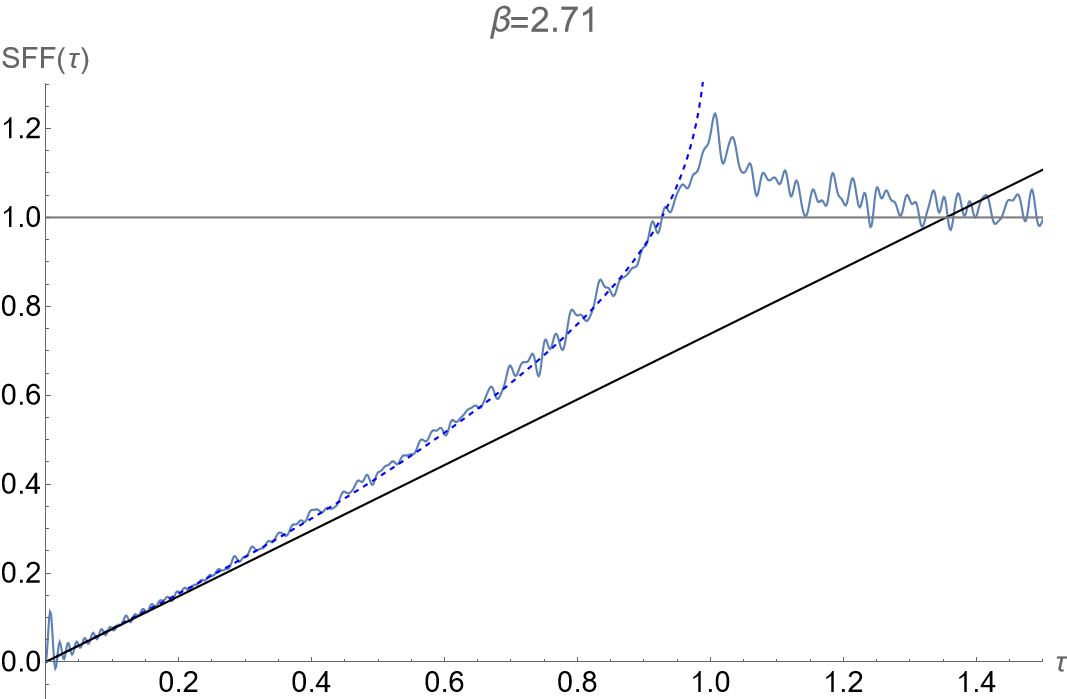} 
   \includegraphics[width=0.48\textwidth]{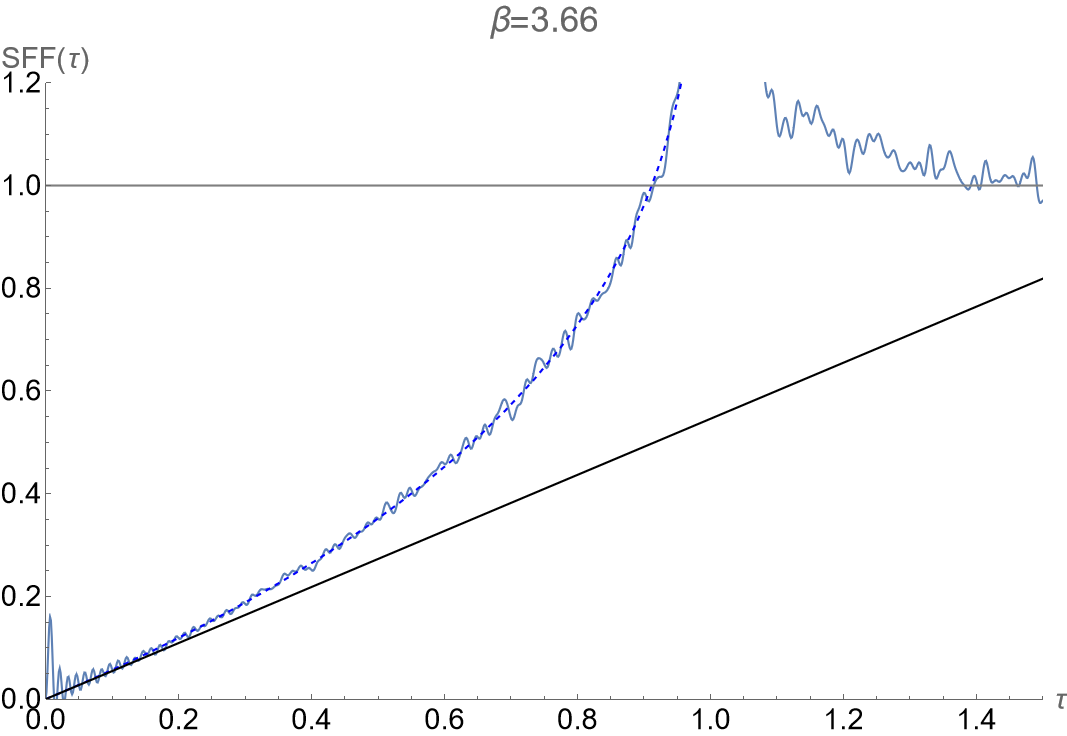} \\
   \caption{Connected part of the SFF $(1-r_2(\tau))$ for $\beta$ between 1 and 4, with our predicted interpolations (dashed), and a linear ramp $\frac{2\tau}{\beta}$ (black line), seen to match at early times. The SFF is computed as an ensemble average over 2000 matrices of size $L=80$.}
\end{figure} 

% \subsection{When $\beta\to0$...}
% [**It could be interesting to show how the SFF behaves for small $\beta$, interpolating from GOE to Poisson at $\beta=0$. If it doesn;t take too long to get results, I will add it.**]

%\clearpage

\section{From SFF to ScFF and their general properties} \label{sec:generalproperties}
In section \ref{sec:beta} we saw explicit expressions for the spectral form factor in the $\beta$-ensembles of RMT. Here we discuss the SFF and its analog, the scattering form factor (ScFF), in more general and physical terms.

\subsection{Spectral form factor for Hamiltonian systems}
Give some Hamiltonian system $H$ of dimension $L$, with energy eigenvalues $E_n$, $n=1,2,\ldots,L$, the partition function is\footnote{Let us recall that the symbol $\beta$ without a tilde is reserved in this paper for the parameter of RMT ensembles, so it is avoided here.}
\be
Z(\tilde\beta)= \Tr{e^{-\tilde\beta H}} = \sum_{n=1}^L e^{-\tilde\beta E_n}
\ee
One can analytically continue this quantity by taking $\tilde\beta\rightarrow \tilde\beta+i t$ and defining 
\be
Z(\tilde\beta,t)= \Tr{e^{-\tilde\beta H-iHt}}
\ee
Here $\tilde\beta$ is the inverse temperature and $t$ is time.

We are interested in the associated function given by
\be\label{therSFF}
{Z(\tilde\beta,t)Z(\tilde\beta,t)^{*}\over Z(\tilde\beta,t{=}0)^{2}}=\frac{1}{|Z(\tilde\beta)|^{2}}\sum_{n,m}e^{-\tilde\beta (E_{n}+E_{m})}e^{it (E_{m}-E_{n})}
\ee
We will define the spectral form factor as a function of time, from the large temperature limit ($\tilde\beta=0$) of the above expression:
\be
\sff(t) \equiv {Z(0,t)Z(0,t)^{*}\over |Z(0,0)|^{2}}= \frac{1}{L^2} \sum_{n,m}e^{it (E_{m}-E_{n})}
\ee

Suppose that the Hamiltonian in question is one of a set of Hamiltonians $\{H^{(I)}\}$, with $I=1,2,\ldots,N_{H}$. This set could be a finite sample of Hamiltonians drawn from a random \emph{ensemble}, defined by some probability function $P(H)$. In this case, we will define the ensemble averaged spectral form factor as a simple average over the sample:
\be\label{enav}
\avg{\sff(t)}=\Big\langle {Z(0,t)Z(0,t)^{*}\over Z(0,t{=}0)^{2}} \Big\rangle_{\{H^{(I)}\}}={1\over N_{H}}\sum_{I=1}^{N_{H}}\left(\frac{1}{L^2}\sum_{n,m}e^{it (E^{(I)}_{m}-E^{(I)}_{n})}\right)
\ee

In many pertinent examples of chaotic systems, where the Hamiltonian is thought of as a random matrix, the SFF is a ``self averaging'' quantity. Meaning, that if one computes the SFF for a single Hamiltonian $H^{(I)}$, and provided that the dimension is large enough ($L\gg1$), then by performing a time average of the SFF of $H^{(I)}$ on an appropriate time scale, one can get a good approximation of the ensemble averaged SFF (see \cite{Cotler:2016fpe} for example).

To uncover the universal properties and compare with the RMT expressions presented in section \ref{sec:beta}, one typically has to compensate for the eigenvalue density function $\rho(E)$, which varies from system to system, and expose the random behavior of the fluctuations in the energy level spacings. The eigenvalue density is defined (in the continuum limit) such that $L\rho(E)dE$ counts the number of eigenvalues in the interval $(E,E+dE)$. Using the cumulative density function, which we define as
\be I(E) \equiv L\int_{-\infty}^E \rho(E^\prime)dE^\prime \ee
we can define the \emph{unfolded} spectrum
\be \lambda_n \equiv I(E_n) \ee
In this variable the density of states is constant, since by definition $\rho(E)dE = \frac{1}{L} d\lambda$.\footnote{In our convention $\rho(E)$ is normalized as a probability density function, $\int \rho dE = 1$, hence the explicit factors of $L$ that appear in the expressions.}

% More generally, for a spectrum of eigenvalues $\lambda_1 \leq \lambda_2 \leq \ldots \leq \lambda_L$, we define the spectral form factor (SFF) as
% \be \sff(t) = \frac{1}{L^2}\sum_{i=1}^L \sum_{j=1}^L e^{i(\lambda_i-\lambda_j)t} = \left\vert \,\frac{1}{L}\sum_{i=1}^L e^{i\lambda_i t}\,right\vert^2 \ee

% To compare results produced by different ensembles, we always normalize the eigenvalues such that the average spacing $\delta_i=\lambda_{i+1}-\lambda_i$ is one:
% \be \avg{\delta_i} = 1 \ee
% or more specifically ``unfold'' the spectrum such that the average eigenvalue density is constant. This will be covered in more detail below.

%%%%%%%%%%%%%%%%%%%%%%%%%%%%%
\subsection{The scattering form factor}
Following the analogy between the (chaotic) distributions of the (normalized) ratios of (energy) eigenvalues and the (chaotic) distributions of the (normalized) ratios of extremum points of a scattering amplitude \cite{Bianchi:2022mhs,Bianchi:2023uby} we propose the analogous ScFF (scattering form factor)
\be
\mathrm{ScFF}(s)=\frac{1}{L^2}\sum_{i,j} e^{is(z_i-z_j)}
\ee 
where $z_i$ and $z_j$ are the locations of the extrema of a scattering amplitude, when considered as a function a continuous physical variable, such as the scattering energy, Mandelstam variables, or of the scattering angle. Here $L$ is the number of peaks, and the variable $s$ is taken to be real, such that $\mathrm{ScFF}(s)$ is also a real function, since $z_i$ are real.

Let us consider now the amplitude as a function of a single angle, ${\cal A}(\alpha)$. The extrema can be found by computing the zeros of the logarithmic derivative
\be
 {\cal F}(\alpha)\equiv \frac{d \log{\cal A}}{d \alpha}
\ee
such that the ``eigenvalues'' are now the points solving the equation
\be
{\cal F}(z_{i})=0
\ee

By definition it follows that $\mathrm{ScFF}(s=0)=1$. In  complete analogy to the ``ensemble averaging" \eqref{enav} of the SFF one can also define an ensemble averaging also for the ScFF by averaging over large $N_S$ sets $\{z_i\}$ of eigenvalues, which can be generated for example by considering the scattering of different asymptotic states with different amplitudes ${\cal A}(\alpha)$. In the string theory case we will consider below, the averaging will be done by analyzing a sample of scattering amplitudes for many possible excited states of the string at the same mass level.

Unlike the usual spectral form factor, $s$ is not naturally interpreted as a time variable. Rather, as the conjugate variable of an angle $\alpha$, we can see the ScFF as a function of angular momentum. We will explore this interpretation of $s$ further below.

As with the energy spectra, in order to compare the ScFF with the random matrix theory predictions, it is best to unfold the resulting ``spectrum'' by changing to a variable where the eigenvalue density is constant. This also helps resolve ambiguities in defining the physical variable. For instance, if we look at $2\to2$ scattering process at fixed center of mass energy (fixed Mandelstam $s$), we can analyze the resulting amplitude either as a function of the scattering angle $\theta_s$ or as a function of the Mandelstam variable $t \approx -\frac12s(1-\cos\theta_s)$. In both cases, the unfolding procedure should take us to the same variable.
  %%%%%%%%%%%%%%%%%%%%%%%%%%%%%%%%%
\subsection{Features of the scattering form factor}
\subsubsection{Continuum limit}
In the limit of $L\rightarrow \infty$ the distribution of the $z_i$ can be approximated by a continuous function
\be
z_j= \sqrt{L} {\cal Z}(z) \qquad z=\frac{j}{L}\qquad \frac{1}{L}\sum_j\rightarrow \int \rho(z) dz
\ee
where we have introduced the ``eigenvalue" density 
\be
\rho(z)=\frac{d{\cal Z}}{dz}
\ee
so that the ScFF becomes
\be \mathrm{ScFF}(s) = \frac{1}{L^2}\sum_{i,j}e^{i(z_i-z_j)s} \to \int dz dz^\prime \rho(z)\rho(z^\prime) e^{i(z-z^\prime)s} \ee
%In this limit the average of $\scff$ takes the form 
%\be
%\langleScFF(s)\rangle=\left\vert \int d z \rho_0e^{is z}\right\vert ^2+  \int\int dz d z' \langle \rho(z)\rho(z')\rangle %e^{i(z-z') s}
%\ee
%The continuum limit is a good approximation only for small values of $s$.
Let us write
\be
\rho(z)=\rho_0(z) +\delta \rho(z) \qquad \langle\rho(z)\rangle=\rho_0(z)\qquad \langle\delta \rho(z)\rangle=0
\ee
where $\langle ...\rangle$ denotes ensemble average. Then, $\rho_0(z)$ is the \emph{average eigenvalue density} function. Then
\be
\langle\rho(z)\rho(z^\prime)\rangle=\rho_0(z)\rho_0(z^\prime)+\langle\delta\rho(z)\delta\rho(z^\prime)\rangle
\ee
and we get that 
\be
\langle \mathrm{ScFF}(s)\rangle=\left\vert{\int d z \rho_0(z) e^{is z}}\right\vert^2+ \int dz d z^\prime \langle\delta \rho(z)\delta\rho(z^\prime)\rangle e^{i(z-z^\prime) s}
\ee
The first term is the \emph{disconnected} part, which is given by the Fourier transform of the average density function $\rho_0(z)$. The second term is the \emph{connected} part, which depends only on the fluctuations around the average density.
%%%%%%%%%%%%%%%%%%%%%%%%%%%%%%%%%%%

\subsubsection{Small $s$ behavior: Decline}
In the region of small values of $s$, which is referred to as the slope \cite{Cotler:2016fpe} or the decline, the main contribution to $SFF(s)$ comes from $SFF_d(s)$.
\be
ScFF(s)_{disc}=|Z(s)|^2 \quad , \quad  Z(s) =\frac{1}{L}\sum_{z_i} e^{isz_i}
\ee  
The decline region  of the scattering ScFF in a similar manner to the spectral one,  is a result of the  bounded eigenvalue density. The precise power-law is determined by the edge of the density function.

For  a density  $\rho_0(z)$ which takes the form of the Wigner semicircle law, 
\be
\rho_0(z)=\frac{1}{2\pi}\sqrt{4-z^2}
\ee
the ensemble average of the partition function  is given by 
\be
\langle Z(s)\rangle_{SC}= \int_{-2}^2 dz L \rho_0(z) e^{-i s z} = L \frac{J_1(2s)}{s}
\ee
where $J_1$ is a Bessel function and we introduced a scale $L$. Therefore  the ScFF decays for large values of $s$ in  the decline region\footnote{Large refers to values of $s$ such that the asymptotic expansion of the Bessel function holds true, yet small with respect to the `dip' (at $s\sim \sqrt{L}$) and the `plateau' (from $s \sim L $), see below.} as
\be
 ScFF(s) \sim\frac{1}{s^3}.
\ee
until $ScFF(s)$ reaches a minimum called the `dip'.
If on the other hand we take a constant density  $\rho_0(z)$ we get\cite{Delacretaz:2022ojg}
\be
z_i(s) \simeq\frac{\int_{z_{i_{min}}}^{z_{i_{max}}} d z e^{isz}}{z_{max}-z_{min}}=\frac{e^{isz_{max}}-e^{isz_{max}}}{i(z_{max}-z_{min})s}
\ee
and then the decay of the  ScFF in the decline region is according to  
\be
ScFF(s)\sim \frac{1}{s^2}
\ee

Note that there are two scales at play here. We are discussing small $s$, yet the decay of $s^{-2}$ or $s^{-3}$ is in the range where $\scff_d(s)$ can be expanded for large argument. This regions is seen at small $s$ compared with the characteristic scale introduced by $L$.

\subsubsection{Plateau}
In the large $s$ domain, we start with the region of very large $s$. The average of $\scff(s)$ in the limit of large $s$ is given by  
\be
lim_{S\rightarrow \infty} \frac{1}{S}\int_0^S ds ScFF(s)
\ee
If we assume no degeneracies of the various values of $z_i$ and we assume that only $z_i=z_j$ contribute to the sum, due to cancellations between random phases, it is obvious that 
\be
lim_{S\rightarrow \infty} \frac{1}{S}\int_0^S ScFF(s)=\frac{1}{L}
\ee
This is a universal behavior that follows from   the discreteness and finiteness of the spectrum of the $z_i$ eigenvalues.

\subsubsection{Ramp}
An important characteristic of chaotic behavior is the {\it Ramp} which describes evolution of the form factor in 
the region that connects the decline at small values of $s$ to the plateau  at large values of $s$. The RMFF, the spectral form factor of RMT, admits a ramp and so do systems with chaotic spectra using the measure of the ratio of the spacings. Thus, one anticipates that also the scattering form factor associated with chaotic scattering amplitudes will have a ramp region in between the decline and the plateau.

Let us briefly review what a ramp behavior of the ScFF means in analogy to that of the RMT.\footnote{Here we follow \cite{Cotler:2016fpe}}. 

The connected ScFF can be written as 
\be
ScFF_c(s) = \int dz_1 dz_2 R_2(z_1,z_2) e^{is(z_1-z_2)} \qquad  R_2(z_1,z_2)\equiv{\langle\delta\rho(z_1)\delta\rho(z_2)\rangle}
\ee
where $R_2(z_1,z_2)$  is the connected pair correlation function of the density of eigenvalues.
In the RMT near the center of the semi-cicle  $R_2(z_1,z_2)$ is given by a negative {\it sine kernel} on top of a delta function at coincident points.
\be
R_2(z_1,z_2)= -\frac{sin^2[L(z_1-z_2)]}{[\pi L ( z_1-z_2)]^2} \frac{1}{\pi L}+ \delta(z_1-z_2)
\ee
Inserting this into the expression for the connected ScFF we get
\be
ScFF_c(s) \sim \begin{cases} \frac{s}{2\pi L^2} & s< 2L\\ \frac{1}{\pi L} & s\geq 2L \end{cases}
\ee
Comparing this result to the one in the decline we conclude that values of $s$ at the dip and the plateau points are approximately  given by
\be 
s_d\sim \sqrt{L} \qquad  s_p\sim L
\ee
%is called the . We use the expansion of $\rho(z)$ to second order in the fluctuations and insert it to the expression of the action of the fluctuations. Using it one finds 
%\be
%ScFF_c(s)=\frac{|s|}{n^2 \beta \pi}
%\ee
%This implies that the slop of the ramp is inversely proportional to $\beta$ and to $n^2$.

We end with defining the notion of  {\it Scattering rigidity } which is the analog of the spectral rigidity \cite{Dyson:1962es}.
Expanding the relevant action to quadratic order around the (large $N$) saddle point one expects to find  
\be
\delta I= -L^2\int dz dz' \rho(z)\rho(z') \log|z-z'|
\ee
Fourier transforming the perturbations of the `eigenvalue' density
\be
\delta \rho(z) = \int d k \widetilde{\delta \rho}(k) e^{i k z}
\ee one gets
\be
\delta I = \frac{L^2}{2} \int dk \widetilde{\delta \rho}(k)\frac{1}{|k|}\widetilde{\delta \rho}(-k)
\ee
which implies  that the long-wavelength 
fluctuations of $\rho$ are strongly suppressed. This is the  {\it spectral
rigidity} of RMT and we expect that for chaotic scattering a similar behavior will take place for the scattering rigidity. 
%%%%%%%%%%%%%%%%%%%%%%

%%%%%%%%%%%%%%%%%%%%%%%%%%
\subsubsection{Thermal version of the ScFF}
The basic spectral form factor defined in (\ref{therSFF}) is a function of time and the temperature of the system. In a similar manner one can define a ``thermal" generalization of the scattering form factor by elevating $s\in{\mathbb{R}}$ into a complex variable $s-i\tilde\beta$. We can define the partition function
\be
Z(\tilde\beta,s)=\sum_{i=1}^L e^{(i s-\tilde\beta)z_i}
\ee
and the generalized ScFF
\be\label{ThermVers}
\mathrm{ScFF}(\tilde\beta,s)=\frac{|Z(\tilde\beta,s)|^2}{Z(\tilde\beta)^2} =\frac{1}{Z(\tilde\beta)^2}\sum_{i,j} e^{-\tilde\beta(z_{i}+z_{j})}e^{is (z_i-z_j)}
\ee 
where $\tilde\beta$ is the inverse of the ``scattering temperature''.

If we are considering an amplitude's angular dependence, and $s$ is interpreted as angular momentum, then $\tilde\beta$ corresponds to a complexification of the angular momentum. This is a common notion in studying scattering amplitudes, specifically in Regge theory.

\subsubsection{Symmetry resolved ScFF}
In \cite{Delacretaz:2022ojg} a variant of the spectral form factor was defined related to the time  reversal invariance of the $\phi^4$ theory.
Our scattering amplitude is invariant under $\theta\leftrightarrow (\pi-\theta)$. 
Thus we propose to define the symmetry-resolved scattering form factors 
\be
ScFF_{++}= Z_{+}(s) Z^*_{+}(s)\quad ScFF_{--}= Z_{{-}}(s) Z^*_{-}(s)
\ee
and 
\be
ScFF_{+-}= \frac12(Z_{+}(s) Z^*_{-}(s)+Z_{-}(s) Z^*_{+}(s) )
\ee
where 
\be
Z_+(s)= \frac{1}{N_+}\sum_{i+} e^{iz_{i+}s}\,, \qquad Z_-(s)= \frac{1}{N_-}\sum_{i-}e^{iz_{i-}s}
\ee
are the partition functions in the even and odd sectors respectively.
These form factors  provide additional information on the chaotic nature of  scattering amplitudes. 

%-------------------------------------------------

\section{SFF and ScFF for non-chaotic systems}
For the sake of comparison and in order to acquire some familiarity with the spectral form factor 
\be\label{gsscff}
g(t,\tilde\beta) = \bigg|\frac{Z(\tilde\beta, t)}{ Z(\tilde\beta)}\bigg|^2 
\ee
in some elementary context, we study two extremely simple (almost trivial) integrable systems: the harmonic oscillator and the free open string. At the end, we will pass to consider the scattering form factor for the (non-chaotic) 2-body decay of a massive higher spin particle into scalars in order to gain some insights in the possible interpretation of the $s$ and of its complex "thermal" generalization to $s^\prime=s+i\tilde\beta$. In the next section we will discuss in some detail the first example of a scattering form factor for a chaotic process, the scattering in the leaky torus.

\subsection{Harmonic oscillator} 
The free harmonic oscillator represents one simple case of non-chaotic system since it is well known that there is a trivial correlation between the eigenvalues $E_n=\omega(n+{1\over 2})$. 
The respective SFF can be computed starting from the partition function of the system 
\be
Z(\tilde\beta, t) = \sum_{n=0}^\infty e^{(it-\tilde\beta)\omega (n+{1\over 2})} = {1 \over 2 \sinh {(\tilde\beta-it)\omega\over 2}}
\ee
and plugging it in the expression \eqref{ThermVers}, with the result
\be\label{Harmosc}
g(t,\tilde\beta) = \left\vert {Z(\tilde\beta, t)\over Z(\tilde\beta)}\right\vert^2 = \left\vert {\sinh {\tilde\beta\omega\over 2}\over \sinh {(\tilde\beta-it)\omega\over 2} }\right\vert^2 = {1\over \cos^2{\omega t\over 2} + \sin^2{\omega t\over 2} \coth^2{\omega \tilde\beta\over 2} }
\ee
Following the normalization one can see that $g(t{=}0, \tilde\beta){=} 1$ while for asymptotically large $t$ the SFF oscillates between $1$ for $\omega t = 2\pi n$ and $1/\coth^2(\tilde\beta\omega/2)$ for $\omega t = (2n+1)\pi $, on average in a (half) period
\be
\langle g(t,\tilde\beta)\rangle_T = {\omega\over \pi} \int_0^{\pi\over \omega} {dt \over 
\cos^2{\omega t\over 2} + \sin^2{\omega t\over 2} \coth^2{\omega \tilde\beta\over 2} }
={1\over  \coth(\tilde\beta\omega/2)}
\ee
which is clearly extremely small for $\tilde\beta\rightarrow 0$ (high temperature) and yields $1$ for $\tilde\beta\rightarrow \infty$ (low temperature).

A variant of this (trivial) analysis is to set $\tilde\beta=0$ to start with and ``regulate`' the sum (partition function) by including only the first $N$ levels and defining
\be
g(t,\tilde\beta) ={1\over N^2} \left\vert Z_N(\tilde\beta=0, t)\right\vert^2 =  {1\over N^2} \left\vert\sum_{n=0}^{N-1} e^{-in\omega t}\right\vert^2 = {1\over N^2} {\sin^2 {N\omega t \over 2} \over \sin^2 {\omega t \over 2}}
\ee
that oscillates between $1$ (for $\omega t = 2n\pi$) and $0$ for $\omega t = {2n'\pi \over N}$ with $n'\neq kN$.

As expected there is no sign of the typical features of a chaotic system. On the other hand, by introducing by hand some ``noise'' to the spectrum of the harmonic oscillator spectrum, the authors of \cite{Das:2023yfj} were able to argue for the onset of some form of chaotic behaviour even in this elementary model.

\subsection{Free open strings}
Another perennially interesting example of a non-chaotic system we want to discuss is the free open bosonic string\footnote{Without loss of generality we study the open bosonic string spectrum without considering the center of mass zero modes and factoring out Chan-Paton factors since they are irrelevant in the present analysis.},
with spectrum $\lambda_n = n-1$ in units of $\alpha^\prime{=}1$.

The partition function of the system in $d$-dimensions is coded in 
\be
Z(\tilde\beta) = \prod_{n=1}^\infty {1\over (1- e^{-\tilde\beta n})^d} = {e^{- {\tilde\beta d \over 24}} \over \eta_{_{D}}(q{=}e^{-\tilde\beta})^d}
\ee
where $\eta_{_{D}}(q)$ is Dedekind eta function. The SFF can be computed from (\ref{ThermVers}). AS a result one gets 
\be\label{sffstring}
g(t, \tilde\beta) = \left\vert {Z(\tilde\beta+it) \over Z(\tilde\beta)}\right\vert^2 = \left\vert {\eta_{_{D}}(q)^d \over \eta_{_D}(e^{it} q)^d}\right\vert^2
\ee
where the variable $q$ now can be expressed as
\be
\tilde{q} = e^{2\pi i \tau} = e^{2\pi i (\tau_1 + i\tau_2)} = e^{2\pi i \tau_1} q
\ee
In figure \ref{etaplot} one can see the plot of the SFF inside a period $t\in[0,2\pi]$ for various values of the inverse temperature $\tilde\beta$. 
\begin{figure}[h!]
    \centering
    \includegraphics[scale=0.45]{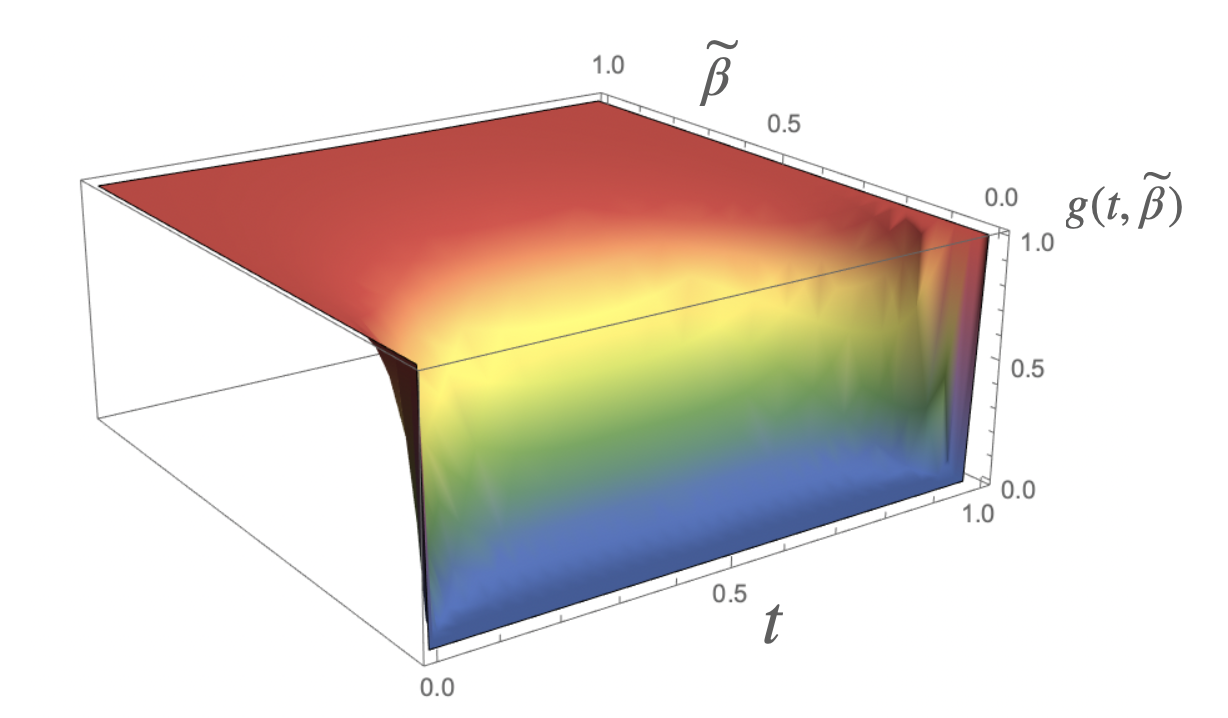}
    \caption{3D Plot of the SFF of a one dimensional free open string according to \ref{sffstring}.}
    \label{etaplot}
\end{figure}
\begin{figure}[h!]
    \centering
    \includegraphics[scale=0.4]{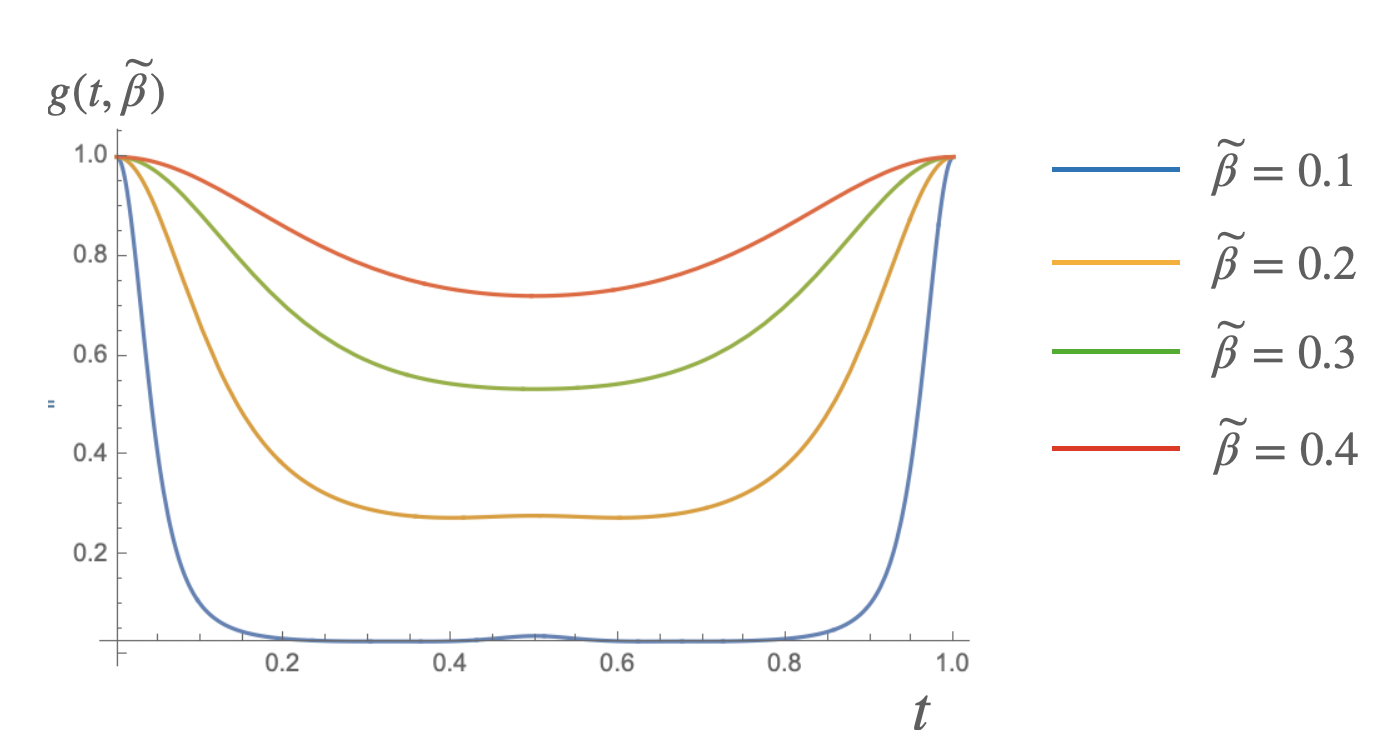}
    \caption{SFF of a one dimensional free open string for fixed values of $\widetilde\beta$.}
    \label{etaplot}
\end{figure}

Proceeding by analogy with the previous system, one can make the following identifications
\be 
2\pi \tau_1 = t\,, \quad 2\pi\tau_2 = \tilde\beta 
\ee
which translates, periodicity of the SFF under $t\rightarrow t+ 2\pi$, and in general the modular properties under 
\be
\tau \rightarrow \tau +1 \,,\quad \tau \rightarrow -1/\tau\,;\quad \tau=\tau_1+i\tau_2={t+i\tilde\beta\over 2\pi}
\ee
reproducing the analog of the temperature inversion 
\be
\tilde\beta\rightarrow 1/\tilde\beta\,,\quad t\rightarrow t/\tilde\beta
\ee
Notwithstanding the presence of many oscillations, yet with tiny amplitude, the SFF of the free string (\ref{sffstring}) does not expose any of the desired features of a chaotic Hamiltonian system. In fact the regularity of the oscillations of (\ref{sffstring}) at special rational values of the argument will have a consequence (an additional ``bump'' structure) on the chaotic behavior of the ScFF for the decay of a HESS into 2 tachyons that we will analyze later on.

\subsection{Decay of leading Regge trajectory states into two tachyons: Legendre spectrum }
As a warm up for the decay process of a generic highly excited string state into two tachyons, let us consider the ScFF in a simple 2-body decay model in $D=3+1$ dimensions.

Parametrizing the two-body decay amplitude of a spin $\ell$ boson of mass $M_H$ into 2 scalars of mass $M_0$ as
\be
A_\ell(H; p_1,p_2) = g_\ell H_{\mu_1 \ldots \mu_\ell} (p_1-p_2)^{\mu_1} \ldots (p_1-p_2)^{\mu_\ell}
\ee
with $g_\ell$ a dimensionful constant and $H_{\mu_1\ldots \mu_\ell}$ a totally symmetric, transverse (w.r.t. to $P=p_1+p_2$) and traceless tensor.

In the rest frame of the decaying boson
\be
P=(M,0)\,,\quad p_1=(M_H/2, p/2)\,,\quad p_2=(M_H/2,-p/2)
\ee
with
$p^2= M_H^2 - 4M_0^2$, the
$H$ tensor has only space-like components $H_{i_1...i_\ell}$ and is
symmetric traceless.
In terms of the unit vector $u^i = p^i/p$ one can expand the amplitude in the basis of spherical harmonic functions 
\be
H_{i_1...i_\ell}u^{i_1} ...u^{i_\ell} = \sum_{m=-\ell}^{+\ell} C_m Y_{\ell,m} (\theta_u, \phi_u)
\ee
with the coefficients $C_m$ depending on the chosen `polarization' $H$.
Focusing for istance on the $m=0$ component, the relevant spherical harmonic reduces to
$Y_{\ell,0}(\theta_u, \phi_u) = P_\ell(\cos\theta_u)$
and thinking of a highly excited state, one can study the large $\ell$ asymptotic behavior of such $m=0$ component, which turns out to be 
\be
P_\ell(\cos\theta) \approx \sqrt{\theta\over \sin\theta} J_0\left((\ell+{1\over 2})\theta\right) \approx \sqrt{2\over \pi\sin\theta} \cos\left((\ell+{1\over 2})\theta -{\pi\over 4}\right)  \ee
Barring the spurious singularity at $\theta =0$, which lies outside the region of valdity of the approximation, since $P_n(1=\cos(0)) = 1$ as suggested by the first equality, the zeros are given by
\be
\theta_k = \pi {k+{1\over 4}  \over \ell+{1\over 2}}
\ee
with $k=1,... \ell$ for $\theta$ in a single `period' 
 $(0,\pi)$. Now, introducing $z = \theta/\pi \in (-1,+1)$ as
the analogue of the eigenvalues in the scattering context, one can apply (\ref{gsscff}) and compute the scattering form factor to find 
\be
ScFF(s) = {1\over \ell^2} \sum^{1,\ell}_{i,j} e^{is (z_i - z_j)} = \left\vert{1\over \ell} \sum_{k=1}^\ell 
\exp\left({is { k+1/4 \over \ell+{1/2}}}\right)\right\vert^2 = {\sin^2{\ell s\over 2(\ell+1/2)} \over \ell^2 \  \sin^2{s\over 2(\ell+1/2)}} 
\ee
The result is compatible with the SFF of the Harmonic Oscillator (\ref{Harmosc}), in fact with the identification 
\be
\omega t = s/2(\ell+1/2)
\ee
one can see the precise correspondence of the HESS scattering form factor with the spectral form factor of the harmonic oscillator.
%$ScFF_\ell(s)$ is 1 for $s=0$, oscillates indefinitely with period $ (2\ell+1)\pi $, has zeros at $s= (2 + (1/\ell))k\pi ... $ NO ramp, no plateau, NO slope (or maybe many slopes and ramps ... and maybe an average plateau after integrating over $s$ ).
This specific case is illustrating a non chaotic scattering of special HESS i.e. those with $\ell=N$ (first Regge trajectory) whose maximal helicity component corresponds to $J=\ell$ and minimal helicity to $J=0$. The zero's are regularly spaced and the distribution of the (normalized) spacings $\delta_k = C(z_{k+1} - z_k)$ is peaked at $\delta=1$, whether one has a single zero, like for maximal helicity $J=\ell$ whereby $Y_{\ell,\pm\ell}(\theta, \phi=0) = (\sin\theta)^\ell$, or many like for zero helicity $J=0$ whereby $Y_{\ell,0}(\theta, \phi) = P_\ell(\cos\theta)$ is independent of $\phi$.

 The lesson we would like to learn from this elementary model of 2-body decay of massive higher spin boson concerns the interpretation of $s$. Taking (logarithmic) derivatives with respect to $s$ or its complexification $s'=s+i\tilde\beta$ one expects to find some average $z=\theta/\pi$. Given the semi-classical (eikonal) relation 
\be
{\partial \delta (E,J) \over \partial J} \approx {\partial S \over \partial J} = \Delta\theta
\ee
it is natural to think of $s$ as some angular momentum or partial wave number that becomes continuous in the large spin limit of a process. This should hold true in the chaotic process involving highly excited string states with many different spin components which we will discuss in section \ref{sec:stringsff}, as in the simple model with a single (possibly large) spin with several possible projections that we just discussed.

%-----------------------------------------------
%%\clearpage
\section{ScFF of leaky torus as SFF of non-trivial zeroes of Riemann zeta function} \label{sec:leaky}
An extremely interesting model that displays chaotic behaviour is scattering on the leaky torus. Originally proposed by Gutzwiller \cite{Gutzwiller:1983}, the leaky torus geometry is constructed by taking the two dimensional hyperbolic space with the metric 
\begin{equation} ds^2 = \frac{dx^2+dy^2}{y^2} \label{eq:leaky_metric} \end{equation}
where we set radius to unity. One looks at the region, in the upper half plane \(y>0\), between the geodesics (i) \(x=-1\), (ii) \(x = 1\), (iii) \((x-\frac12)^2+y^2 = (\frac12)^2\), and (iv) \((x+\frac12)^2+y^2 = (\frac12)^2\).
Then, identifying boundary (i) with (iii) and (ii) with (iv), the result is a torus with a cusp point at infinity.

The scattering in this setting involves sending an incoming free wave from \(y=\infty\) and measuring the phase shift of the outgoing wave at some finite \(y = y_0 > 0\). The phase shift is found to be exactly
\be S(k) \equiv e^{i\delta(k)} = \frac{\pi^{-ik}\Gamma(\frac12+ik)\zeta(1+2ik)}{\pi^{+ik}\Gamma(\frac12-ik)\zeta(1-2ik)} \ee
where $k = \sqrt{2E}$ is the momentum of the incoming wave.\footnote{The standard convention for partial waves is $S_\ell(E) = e^{2i\delta_\ell(E)}$. For consistency with the original references we absorb the factor of 2 into $\delta(k)$.} 

The Wigner time delay function is given in general by the determinant of the logarithmic derivative of the $S$-matrix, which here reduces to\footnote{We mostly follow chapter 8 of \cite{Hurt:1997quantum}.}
\be T(k) \equiv \frac{d\delta(E)}{dE} = \frac{1}{k}\frac{d\delta(k)}{dk}\ee
It is given explicitly by $T(k) \equiv \frac{1}{k}\tau(k)$ with
\begin{align}
     \tau(k) = 4\left(\log(2\pi)-1-\frac{\gamma}{2}\right)-\frac{1}{\frac14+k^2} + \sum_{k_n > 0}\frac{1}{(\frac14)^2+k_n^2} + \nonumber\\
     + \sum_{k_n > 0}\left(\frac{1}{(\frac14)^2+(k+k_n)^2}+\frac{1}{(\frac14)^2+(k-k_n)^2}\right) 
\end{align}
where the sum runs over the non-trivial zeroes of the zeta function, $z_n = \frac12 + 2i k_n$. The term of the second line comes from the fluctuating part of $S(k)$ (involving only the zeta function), an it is given as a succession of resonances located at the zeta function zeros $k = k_n$, all having the same residue and width.
\begin{figure}
\centering
\includegraphics[width=0.60\textwidth]{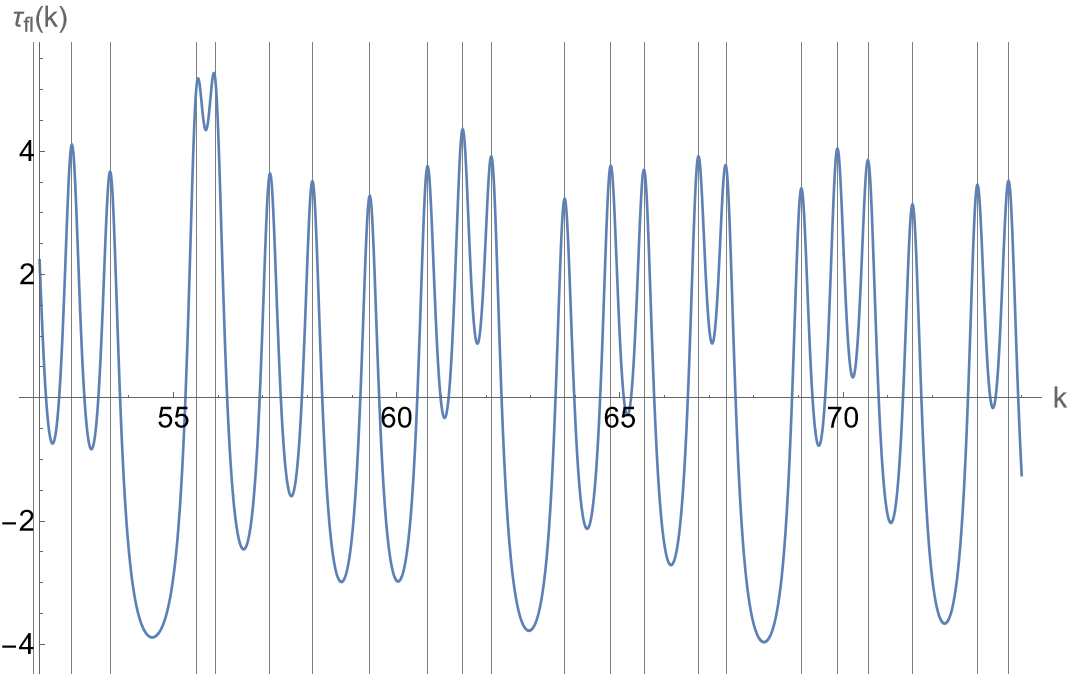} 
\caption{The function $\tau_{\mathrm{fl}}(k)$, displaying erratic behavior as a succession of resonances at zeta function zeros. One can plot any range of $k$ to see similar fluctuations, yet without ever repeating exactly. Vertical lines are located at the zeros of the Riemann zeta, $z_n = \frac12+2ik_n$.}
\label{fig:leaky_function}
\end{figure}

Looking at the fluctuating part of the function,
\be e^{i\delta_{\mathrm{fl}}(k)} \equiv \frac{\zeta(1+2ik)}{\zeta(1-2ik)}\,, \qquad  \tau_{\mathrm{fl}}(k) = \frac{d\delta_{\mathrm{fl}}(k)}{dk} \ee
one can see that it has peaks almost exactly at the zeta function zeros (see figure \ref{fig:leaky_function}), with some small deviations when two neighboring resonances overlap. As a result, the spacings between successive peaks in $\tau(k)$ will follow closely the Wigner-Dyson distribution of the GUE.\footnote{In \cite{Rosenhaus:2020tmv} and our \cite{Bianchi:2022mhs}, the closely related functions $\Phi(k) \equiv \frac{\Imm[\zeta(1+2ik)]}{\Rea[\zeta(1+2ik)]} = \tan\frac{\delta_{\mathrm{fl}}(k)}{2}$ and ${\cal F}(k) \equiv \frac{d\Phi}{dk} = \frac{\tau_{\mathrm{fl}}(k)}{2\cos^2[\frac{\delta_{fl}(k)}{2}]}$ were considered instead of $\tau(k)$ itself. In \cite{Bianchi:2022mhs} we considered the zeros of ${\cal F}(k)$ instead of locations of poles of $\tau(k)$, introducing some deviations from the GUE distribution of spacings of the latter.}

Let us then revisit the distribution of non-trivial zeros of the zeta function, being the solutions of
\be \zeta(\frac12+i y_n) = 0 \ee
To see the distribution of spacings, first we use the average density function, which is known to be
\be \rho(y) = \frac{1}{2\pi} \log\frac{y}{2\pi} \ee
and we define the unfolded spectrum using the cumulative density
\be \lambda_n \equiv I(y_n) = \frac{y_n}{2\pi}\left(\log\frac{y_n}{2 \pi}-1\right) \ee
It is well known that the spacings $\delta_n \equiv \lambda_{n+1}-\lambda_n$ follow the GUE distribution, and as such the zeta function was an object of interest from the early days of quantum chaos \cite{Berry:1986}.

The SFF for the zeta function is intimately related to Montgomery's pair correlation conjecture \cite{Montgomery:1973}, which states that, assuming the Riemann hypothesis to be correct, the two point correlation function for zeros on the critical line is
\be R_2(r) = 1 - \frac{\sin^2(\pi r)}{\pi^2 r^2} \ee
A Fourier transform of this function gives the GUE SFF of eq. \eqref{eq:sff_gue}. In fact, the Montgomery conjecture was the first argument establishing a connection between the zeta function and random matrices.

\begin{figure}\centering
    \includegraphics[width=0.48\textwidth]{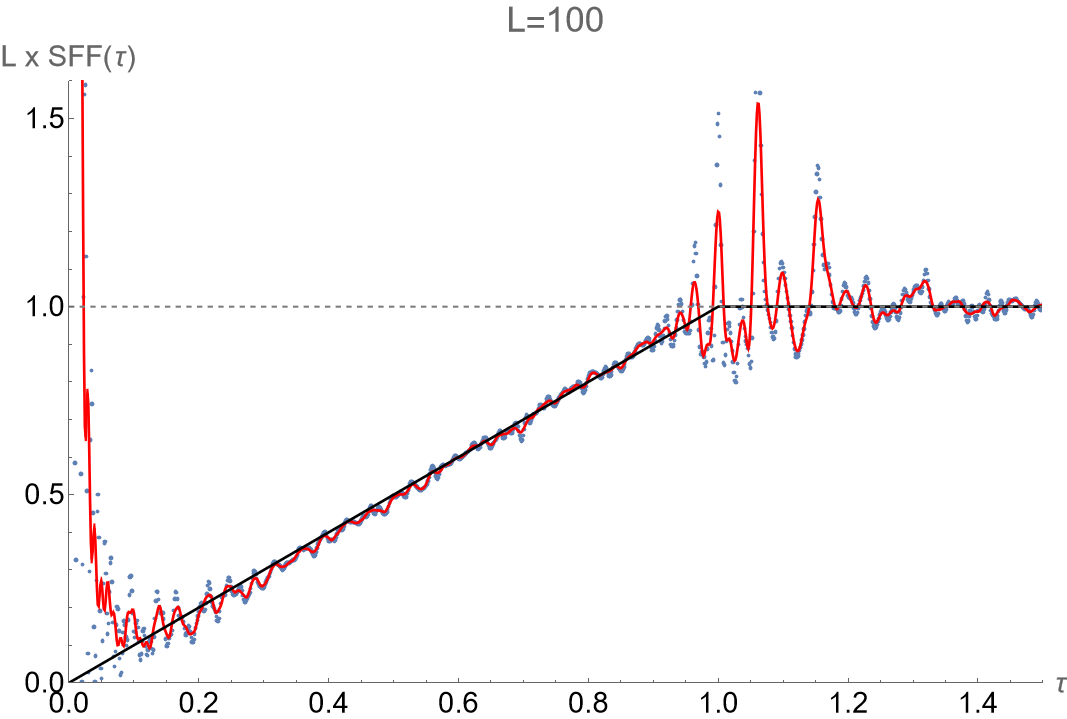}
    \includegraphics[width=0.48\textwidth]{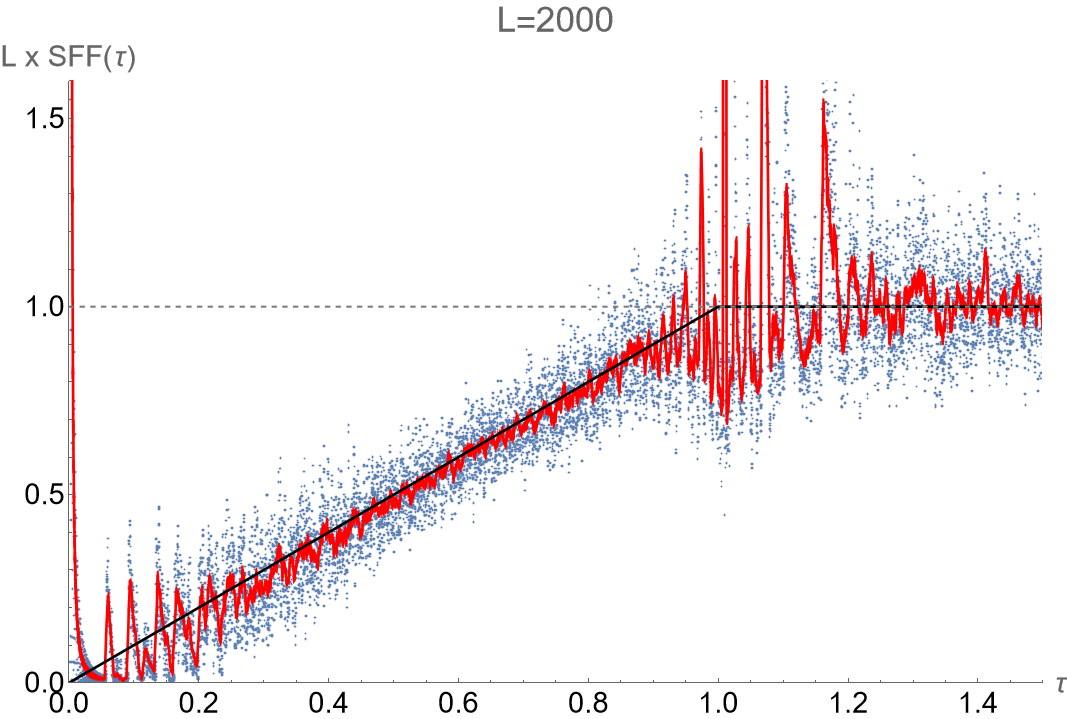}
    \caption{\label{fig:sff_zeta} SFF computed from (unfolded) zeta function zeros. The left figure is an ensemble average over 2000 sets of $L=100$ eigenvalues each. The right is conversely an average over 100 sets of 2000 eigenvalues each. Both match the linear ramp that is the GUE prediction.}
\end{figure}

There are infinitely many zeros of the zeta function. We can plot the SFF for a finite sample of them. To mimic the usual ensemble averaging process, we take the list of the first 2,001,052 zeros given in \cite{Odlyzko:Zeta}, unfold it, split it into sets of $L$ eigenvalues each, and then pick at random a sample of such sets to average over. The result of this process is that the spacings are distributed as in GUE, and the SFF closely follows the GUE prediction (figure \ref{fig:sff_zeta}). There is however a big fluctuation at the end of the linear ramp/the beginning of the plateau section. The most reasonable explanation is that it is an effect of the finiteness of the sample.\footnote{Similar ``eternal'' ramps were seen in \cite{Das:2023yfj}. Once reduced to finite range, they generate wild fluctuations.} 

The ScFF for the leaky torus is seen to be simply that of the GUE, as the SFF of the Riemann zeta function zeros. It is known that the larger class of mathematical $L$-functions generalizing the zeta function have connections to random matrix theory \cite{Conrey:2001}, and it would be interesting to consider them in conjunction with various generalizations of the leaky torus that have already been studied \cite{Hurt:1997quantum}.

%----------------------------------------------------
%%\clearpage
\section{ScFF for string amplitudes} \label{sec:stringsff}
The scattering amplitudes of highly excited string (HES) states in open bosonic string theory were argued in \cite{Gross:2021gsj,Rosenhaus:2021xhm} to display erratic behavior. In \cite{Bianchi:2023uby} we carried out a detailed analysis of the spacing ratios of peaks in these amplitudes and have shown that they display RMT statistics.

In this section, we will continue the analysis of chaos in string scattering amplitudes using the ScFF to further examine the emerging statistics.

Specifically, we will be concerned here with the simplest process involving HES states: the 3-point scattering (2-body decay) amplitude of a single HES state ${\cal H}_{N}^{h}$ and two scalars (tachyons) $\varphi_{1,2}$, which is a function of one angle:
\be
{\cal A}({\cal H}_{N}^{h},\varphi_{1},\varphi_{2}) \equiv  {\cal A}_{N}^{h}(\alpha)
\ee
The chaos is displayed as erratic behavior in this angular dependence.

The HES state is constructed using the formalism of Del Giudice-Di Vecchia-Fubini (DDF) \cite{DelGiudice:1971yjh}. We will briefly review only the key points before writing the form of the amplitude. More details on the DDF method and the computation of the amplitude are found in \cite{Gross:2021gsj,Bianchi:2019ywd}.

In the DDF approach, the HES state is constructed by scattering $h$ photons on a probe tachyon of momentum $\tilde p$. These photons each have a polarization vector $\lambda_i$ and a momentum $n_i q$ with integer $n_i$, and they are on-shell: $q^2=0$ and $\lambda\cdot q = 0$. Formally, one constructs the vertex operator of the HES from the DDF operators,
\be \label{eq:DDF_op}
A_n^i(q) = \oint {dz\over {2\pi}} \partial X^i e^{in qX}
\ee
Then, a general HES state is given by acting with the DDF (creation) operators on the vacuum (tachyon):
\be \vert{\cal H}\rangle = \left(\prod_{i=1}^{h} \lambda_i\cdot A_{-n_i}\right) \vert 0;\tilde p\rangle \label{eq:DDF_state} \ee
We will also take for simplicity all the DDF photons to have the same circular polarization, i.e. $\lambda_i = \lambda$ with $\lambda^2=0$. In this case, the angle $\alpha$ in the HES-tachyon-tachyon amplitude is the one between the momentum of one of the outgoing tachyons and the momentum $q$.

While any excited string state can be written in the form \eqref{eq:DDF_state}, the choice of identical polarizations of the photons restricts us to a specific sector, which nevertheless still represents a large number of ``generic'' string states. The HES states constructed this way can be written as
\be\label{eq:HES_state}
 \vert{\cal H}_N^h\rangle = \prod_{n=1}^N \big(\lambda\cdot A_{-n}(q)\big)^{g_n} |0, \tilde p\rangle
\ee
and are characterized only by a choice $\{g_n\}$ of an \emph{integer partition} of $N$, satisfying
\be
N=\sum_{n=1}^N ng_{n}\,,\quad h = \sum_{n=1}^N g_{n}
\ee
Now $N$ is precisely the \emph{level} of the state, meaning that its mass is $M^2 = (N - 1)/\alpha^\prime$, and $h$ is the \emph{helicity} of the state. 

The amplitude is\footnote{Here we omit constant factors and write only the part that depends on the angle.}
\be {\cal A}_N^h(\alpha) = (\sin \alpha)^h  \prod_{n=1}^{N} \left(\sin(\pi n \cos^2\frac\alpha2)\frac{\Gamma(n \cos^2 \frac\alpha2)\Gamma(n \sin^2 \frac\alpha2)}{\Gamma(n)}\right)^{g_n} \label{eq:A_HTT} \ee

It has been recently established that the distribution of the minima and maxima in this amplitude for a generic partitions of $N$ is chaotic \cite{Gross:2021gsj, Rosenhaus:2021xhm, Bianchi:2023uby, Bianchi:2022qph}. These points are located at the zeros of the logarithmic derivative of the amplitude:
\be
F_{N}^{h}(\alpha)\equiv{d\log{\cal A}_{N}^{h}(\alpha)\over d\alpha }\,, \qquad F_{N}^{h}(\alpha)=0\,\Rightarrow \{\alpha_{k}\}
\ee
The distribution of zeros $\{\alpha_{k}\}$ is connected to the complexity of the HES state. Since all string states can be constructed with DDF operators, it is natural to study the behavior of the zeros $\{\alpha^{(\ell)}_{k}\}$  with respect to the ensemble generated within this approach, including the spin degeneracy of the HES states. 

The number of states, {\it i.e.} the total number of partitions of $N$, is known to asymptote to
\be\label{HESgr19} p_N \approx \frac{1}{4\sqrt{3} N} e^{C \sqrt{N}}\,, \qquad C \equiv \frac{2\pi}{\sqrt 6} \ee
at large \(N\).
The fraction of partitions of fixed length (helicity $h$) is asymptotically given by a Gumbel distribution \cite{Erdos:1941}, i.e. according to the doubly exponential probability distribution
\be d_N(h) = \frac{1}{{\nu}} \exp\left(-\frac{h-\mu}{{\nu}} - e^{-\frac{h-\mu}{{\nu}}}\right) \label{eq:Gumbel} \ee
with the parameters $\mu, \nu$ scaling as \(\mu \sim \sqrt{N} \log N\) and \({\nu} \sim \sqrt{N}\). More precisely the distribution is centered around
\be \langle h \rangle \sim \frac{1}{C} \sqrt{N}\log N \label{eq:typicalJ} \ee
which makes this the ``typical helicity'' of a randomly chosen string state at level \(N\).

So, at level $N$ and fixed $h$, there is a set of amplitudes $\{ {\cal A}_{N}^{h}(\alpha)\}_{d_{N}(h)}$, each with its own ``spectrum'' of zeros $\{\alpha_{k}^{(\ell)}\}_{d_{N}(h)}$. The total number of states grows exponentially in $\sqrt{N}$, and the helicities of the vast majority of the states are in a window of size $\sim \sqrt{N}$ around the typical value of $\sqrt{N}\log N$.

For a state at any fixed $N$ the distribution of the zeros is sensitive to the value of $h$. One can make the following general observations:
\begin{itemize}
\item[a)] $h=1$ 
there is no ensemble because there is only one set of zeros, and the distribution of $\{\alpha_{k}^{(\ell)}\}_{d_{N}(1)}$ is close to the Wigner semicircle with radius $R=1.3$ with the center located at $\pi/2$ 
\be
PDF(\{\alpha_{k}^{(1)}\})\Big|_{h=1}\simeq {\pi\over 2 R^{2}}\sqrt{R^{2}-\left(x-{\pi\over 2}\right)^{2}}
\ee
\item[b)] $h> 1$
the distribution of the zeros deviates from the Wigner semicircle randomly spreading out. 

For example in order to fix the notation one can analyze the first non trivial case with spin degeneracy at level $N=4$, whereby
\be
(N,h) \Rightarrow \{n_{g_{n}}\}\quad \Rightarrow\quad  F_{N=4}^{\{n_{g_{n}}\}}(\alpha)
\ee
The sets of zeros are classified as follows 
\begin{itemize}
\item $h=4$: there is one set of zeros $\{\alpha_{k}^{(\ell)}\}_{d_{4}(4)}$ because $d_{4}(4)=1$ and it is  relative to
\be
F_{N=4}^{h=4}(\alpha)\Rightarrow F_{4}^{(1_{4})}(\alpha) 
\ee 
\item $h=3$ there is one set of zeros, $d_{4}(3)=1$, with  $\{\alpha_{k}^{(\ell)}\}_{d_{4}(3)}$ according to:
\be
F_{N=4}^{h=3}(\alpha)\Rightarrow F_{4}^{(1_{2},2_{1})}(\alpha) 
\ee
\item $h=2$ there are two sets of zeros, $d_{4}(2)=2$, so there is an ensemble of two elements
\be
 \{\alpha_{k}^{(\ell)}\}_{d_{4}(2)}= \Big\{ \{\alpha_{k}^{(1)}\}_{d_{4}(2)},  \{\alpha_{k}^{(2)}\}_{d_{4}(2)}  \Big\}
\ee
according to 
\be
F_{N=4}^{h=2}(\alpha)\Rightarrow \{F_{4}^{(1_{1},3_{1})}(\alpha),\,F_{4}^{(2_{2})}(\alpha) \}
\ee
\item $h=1$ there is one set of zeros, $d_{4}(1)=1$, with  $\{\alpha_{k}^{(\ell)}\}_{d_{4}(1)}$ according to:
\be
F_{N=4}^{h=1}(\alpha)\Rightarrow F_{4}^{(4_{1})}(\alpha) 
\ee
\end{itemize}
\item[c)] $1 < h < N$: Large range where the distribution of zeros is non-trivial. This is the range in which we will focus our following analysis.
\item[d)] $h \approx N$. The single state with $h=N$ is the state on the leading Regge trajectory. There are no zeros in the amplitude and no distribution to speak of. For $h \approx N$, there are only a few zeros per state and the distribution of zeros tends to become trivial.
\end{itemize}

For each of the cases described above, one can define the ``scattering partition function'' associated to any set of zeros $\{\alpha_{k}^{(\ell)}\}_{d_{N}(h)}$ as
\be
Z_{scatt}(\tilde\beta)=Tr_{\alpha_{k}}(e^{-\tilde\beta \widehat{\alpha}})=\sum_{k} e^{-\tilde\beta \alpha_{k}}
\ee
it defines the spectrum of zeros associated to the scattering amplitude.

Finally given the scattering partition function one can introduce the ensemble averaged scattering form factor 
\be
\scff(s)\equiv{1\over d_{N}(h)}\sum_{\ell=1}^{d_{N}(h)}\sum_{n,m}e^{is(\alpha_{n}^{(\ell)}-\alpha_{m}^{(\ell)})}
\ee 
in analogy with (chaotic) Hamiltonian systems.

Since the number of zeros of the given amplitude scales linearly in $N$.  To see their associated distributions, we will need to collect many zeros for different states. We carry out the statistical analysis in two complementary ways:
\begin{enumerate}
    \item By using samples of many different states of intermediate $N$ and perform the ensemble averaging. The samples we take will have many different partitions of $N$, with $h$ always close to its most likely value (maximum of the Gumbel distribution) to represent generic states at that level. The range of $N$ will be from 100, where we begin to see good agreement with RMT, to 1600. With a large enough sample we can perform reliable ensemble averaging. This analysis is similar to the one carried out in \cite{Bianchi:2023uby}.
    \item By using smaller samples of large $N$, where we have several thousand eigenvalues per state and we can begin to see the distributions even for a single amplitude of a any given string state. For this we take $N$ from 10,000 to 40,000, with one or a few states per each $N$.\footnote{All the numbers here were chosen mostly from considerations of computation time, since there is nothing else preventing us from taking larger samples or accessing larger $N$.}
\end{enumerate}

\subsection{Eigenvalue density and unfolding the spectrum} \label{sec:unfolding}
Our previous analysis in \cite{Bianchi:2022mhs,Bianchi:2023uby} focused on the spacing ratios, which are argued to be insensitive to the procedure of unfolding the spectrum. However, for the computation of the ScFF and its comparison with RMT predictions it is preferable to use the unfolded spectrum, and the HTT amplitude has some features that we need to examine more closely for a precise analysis. In particular the eigenvalue density is such that the results are in fact sensitive to the unfolding procedure, which is highly non-trivial for the case at hand.

As a function of $z = \cos^2\frac\alpha2$, the amplitude can be written as
\be {\cal A}(z) = \big(z(1-z)\big)^\frac{h}{2} \prod_{n=1}^N \left(\frac{(1-nz)_{n-1}}{(n-1)!}\right)^{g_n} \ee
To locate the peaks of ${\cal A}(z)$, we study its logarithmic derivative,
\be {\cal F}(z) \equiv \frac{d}{dz}\log {\cal A} \ee
A very simple calculation yields
\be {\cal F}(z) = \frac{h}{2}\left(\frac{1}{z}+\frac{1}{z-1}\right)+\sum_{n=1}^N \sum_{k=1}^{n-1}\frac{g_n}{z-\frac{k}{n}}  \label{eq:FHTT} \ee
which is a superposition of simple poles, located at the zeros of the amplitude. All of the poles have a positive residue, and so for real $z$ the function ${\cal F}(z)$ is monotonously increasing between two subsequent poles. Therefore, between each pair of poles there is a zero of ${\cal F}$.

We can then count the number of zeros by counting the number of poles. This number is equal to the number of \emph{unique} fractions of the form $\frac{k}{n}$, where $n$ is any number that appears in the partition, and $k=1,\ldots n-1$. We count the two poles at $z=0$ and 1 separately. Then the number of zeros is equal to the number of poles minus one. This poses an interesting mathematical counting problem, which we examine in more detail in appendix \ref{sec:numzeros}. We can see that for unconstrained partitions of large $N$, accounting for the statistical properties of integer partitions, the average number of zeros of ${\cal F}$ is approximately given by $0.44 N$. Across different partitions of $N$, the number of zeros is distributed around that value.

\begin{figure}[h!]
   \centering\includegraphics[width=0.48\textwidth]{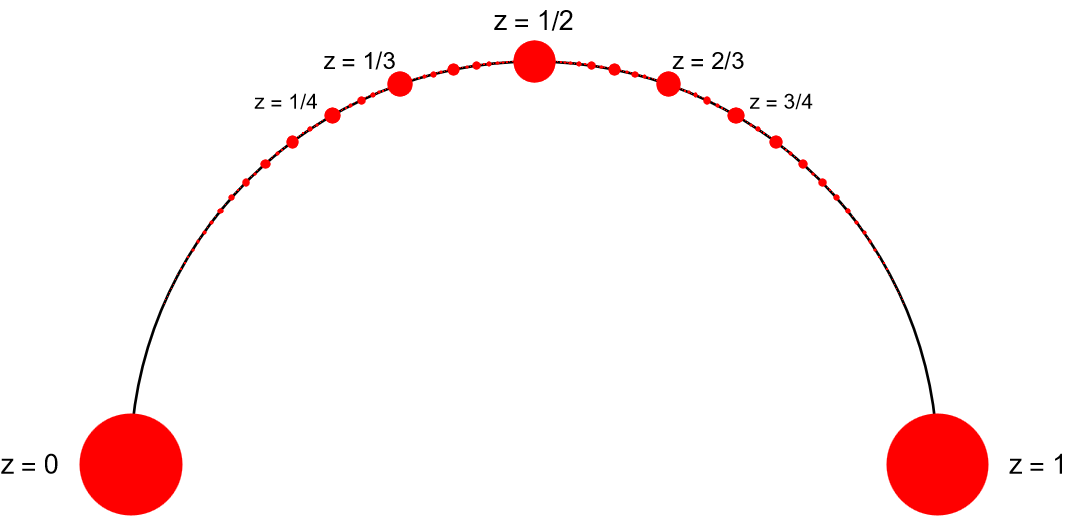} \\
   \includegraphics[width=0.46\textwidth]{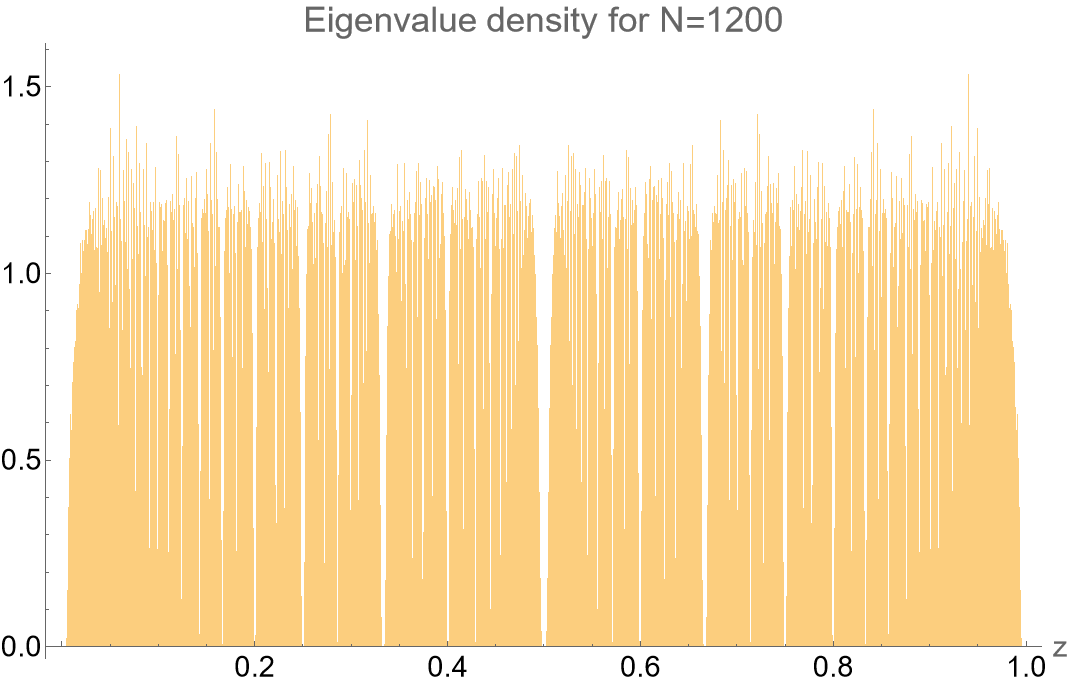}
   \includegraphics[width=0.46\textwidth]{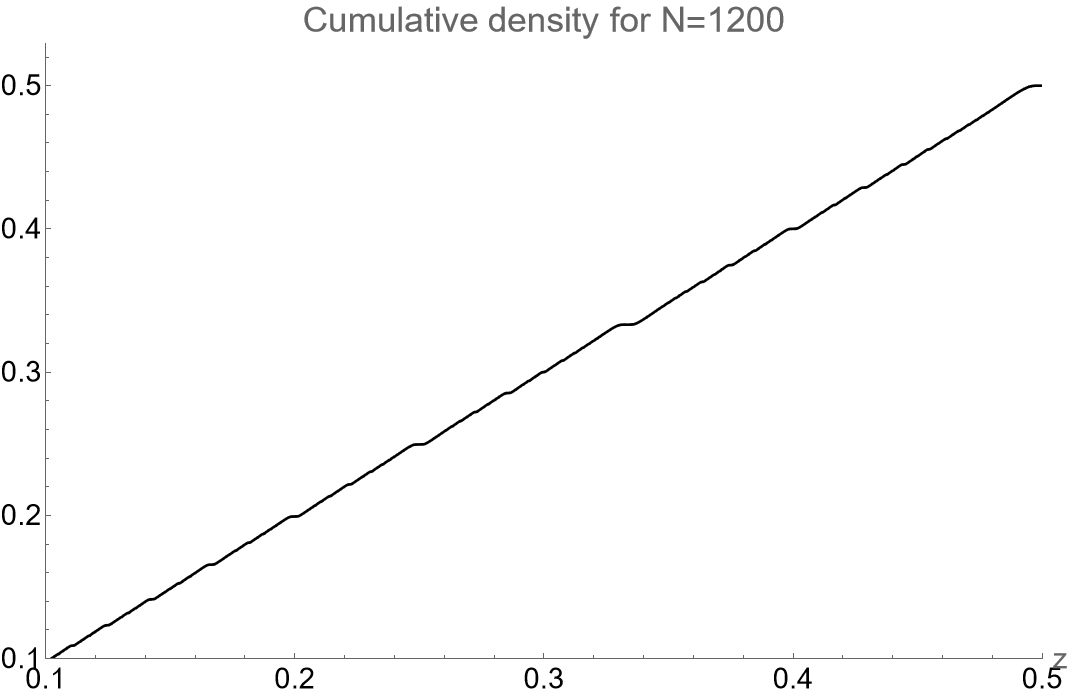} 
   \caption{Top: Eigenvalues are repelled from special points on the circle, where $z = \cos^2\frac{\alpha}{2} = k/n$. The strength of the repulsion depends on the number of times any multiple of $n$ appears in a partition of $N$. The special points are drawn here with the sizes of the circles to scale with the number the points appear as poles in amplitudes for random partitions of $N=1200$. Bottom-left: Histogram for the average density of eigenvalues for a sample of 2000 random partitions of $N=1200$, with visible gaps at the special points. The cumulative density (bottom-right) is a series of constant slopes with flat regions near zeros of $\rho(z)$.}
   \label{fig:eigenvaluedensityHTT}
\end{figure}

The sets $\{z_n\}$ of zeros of ${\cal F}$, which will act as our random matrix eigenvalue spectra, have some extra structure compared with RMT spectra. It is obvious from the form of ${\cal F}(z)$, that the points of the form $z=k/n$ (with $g_n\neq0$) can never be zeros of ${\cal F}$. This means that points like $z=\frac12$, $\frac13$, $\frac14$, $\frac23$, and similar fractions with small denominators are poles for virtually any partition of $N$ that we choose, and consequently can never be zeros. These points can be said to \emph{repel} eigenvalues, as is visible in a plot of the eigenvalue density across a sample of different states (figure \ref{fig:eigenvaluedensityHTT}). The eigenvalue density becomes zero at these special points.

One can choose to focus on sectors in the spectrum away from special points like $\frac12$ and $\frac13$, in the same way as one may try to focus on the center of the spectrum to minimize edge effects in random matrix theory. However, as we increase $N$ more small $n$'s are likely to appear in the partition,\footnote{The probability for any $n$ to appear in a partition of $N$ is known and is written here in appendix \ref{sec:numzeros}.} and the number of repulsive points grows so that we are never free of this phenomenon: all eigenvalues are near some pole.

The usual procedure of unfolding the spectrum can help resolve this issue, or at least minimize its effect. We start with a measurement of the eigenvalue density $\rho(z)$, defined by counting the number of eigenvalues $N(z)$ in some interval $(z,z+\Delta_z)$ and setting $\rho(z)=N(z)/\Delta_z$. Then, we define the cumulative distribution $I(z) = L\int_0^z dz^\prime\rho(z^\prime)$. The presence of zeros in $\rho(z)$ means that $I(z)$ will have some flat regions, which will not be modeled by a simple formula. For instance, if we were to fit $I(z)$ to a polynomial, we would lose this crucial detail and distort the resulting spectra. The best solution is to define $I(z)$ empirically.

Under normal circumstances, that is with a density $\rho(z)$ with no special repulsive points, the distribution of spacing ratios is not sensitive to the unfolding. The simple argument is that if we denote the unfolded spacing ratios by $r_n$ and the ratios of the spacings before unfolding as $r_n^{(0)}$, then typically we could assume that $I(z)$ and $\rho(z)$ change slowly enough on the scale of $\avg{\delta z}$ such that
\be r_n \equiv \frac{I(z_{n+2})-I(z_{n+1})}{I(z_{n+1})-I(z_{n})} \approx \frac{\rho(z_{n+1})(z_{n+2}-z_{n+1})}{\rho(z_{n})(z_{n+1}-z_{n})} \approx r_n^{(0)} \ee
We \textbf{cannot} expect this to work anymore for a density that can go to zero for certain values of $z$.

To further complicate matters, there are many points which appear as pole only in some fraction of the partitions, in which case they will not be totally repellent in the average eigenvalue density. This means that the distribution in each string state would have deviations from the average eigenvalue density. We will see that, when taking a sample of many states at the same $N$, using the measured average eigenvalue density to unfold will introduce some deviations of the resulting spectra from the RMT predictions. In particular we see an unexpected ``bump'' in the ScFF, as will be elaborated in the next section, and a deviation from the Wigner-Dyson distribution for the unfolded spacings, even when the distribution for spacing ratios agrees well with the RMT. These points seem to be resolved when going to very large $N$, where we have enough points per state that we can proceed to measure the eigenvalue density for a single state.

One of the new results of the present analysis is that the distribution of spacing ratios for the string amplitudes is sensitive to unfolding, and that we find a result much closer to the CUE (equivalently, GUE) predictions after unfolding. In \cite{Bianchi:2023uby} it was observed that without unfolding the distribution of spacing ratios for the HTT amplitude is a $\beta$-ensemble distribution (eq. \eqref{eq:beta_r}) with $\beta\to1.7$ for large $N$. Now we find that the unfolding increases the best-fit value of $\beta$ such that the asymptotic value at large $N$ is near the CUE value of $\beta=2$. In the next subsection we will discuss the ScFF's computed from the same eigenvalues, and see that they also suggest a distribution close to CUE with some deviations and new effects.

\begin{figure}[h!] \centering
    \includegraphics[width=0.80\textwidth]{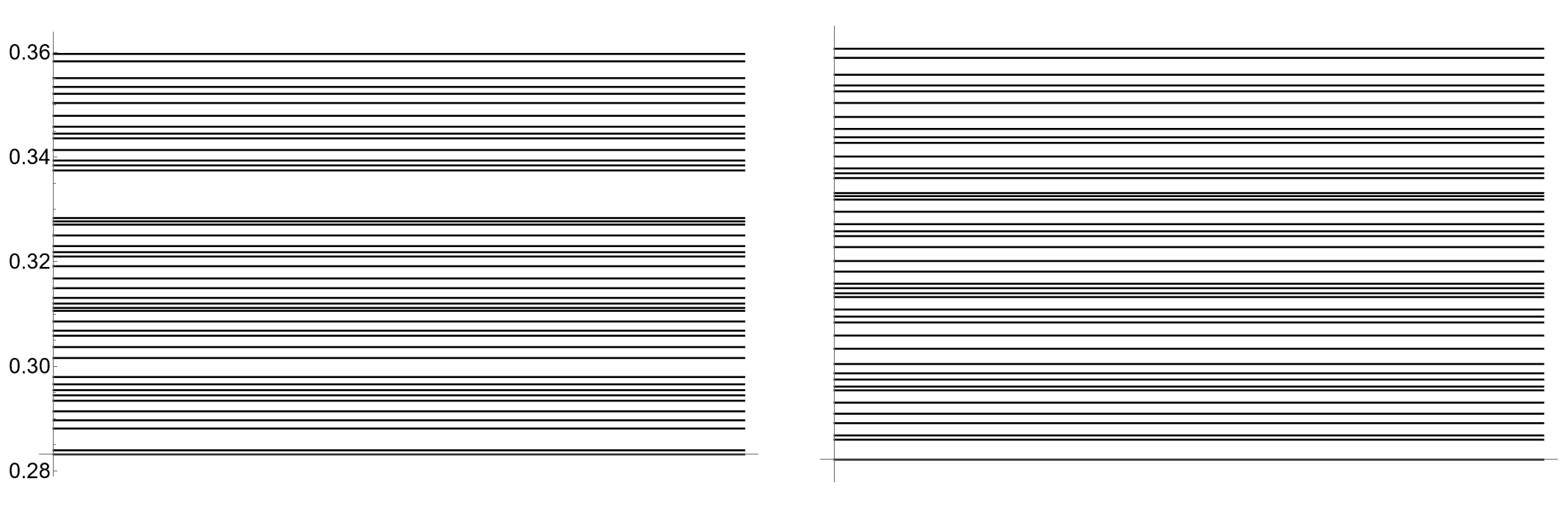}
    \caption{Part of the spectrum for a single state at $N=1200$ before (left) and after (right) unfolding. In particular, the left plot shows a strong repulsion from $z=\frac13$, with smaller gaps at $\frac27$ and $\frac3{10}$. The unfolding has the overall effect of closing the gaps in this distribution to expose the underlying RMT structure of the spacings.}
   \label{fig:unfolding}
\end{figure}

\begin{figure}
    \centering
    \includegraphics[width=0.46\textwidth]{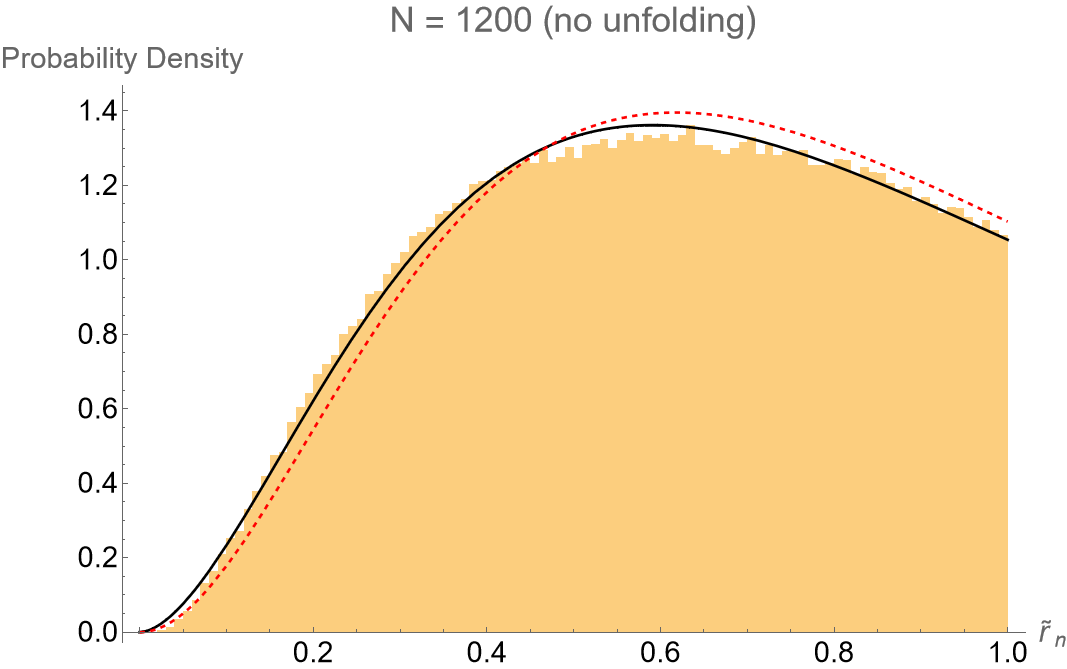}
    \includegraphics[width=0.46\textwidth]{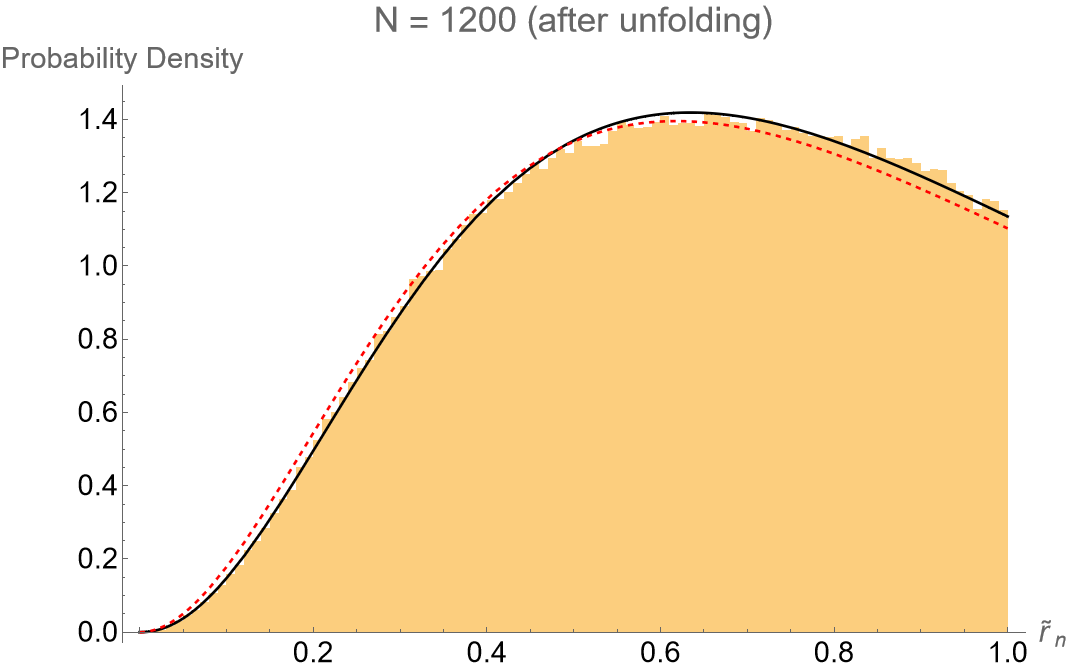}
    \includegraphics[width=0.46\textwidth]{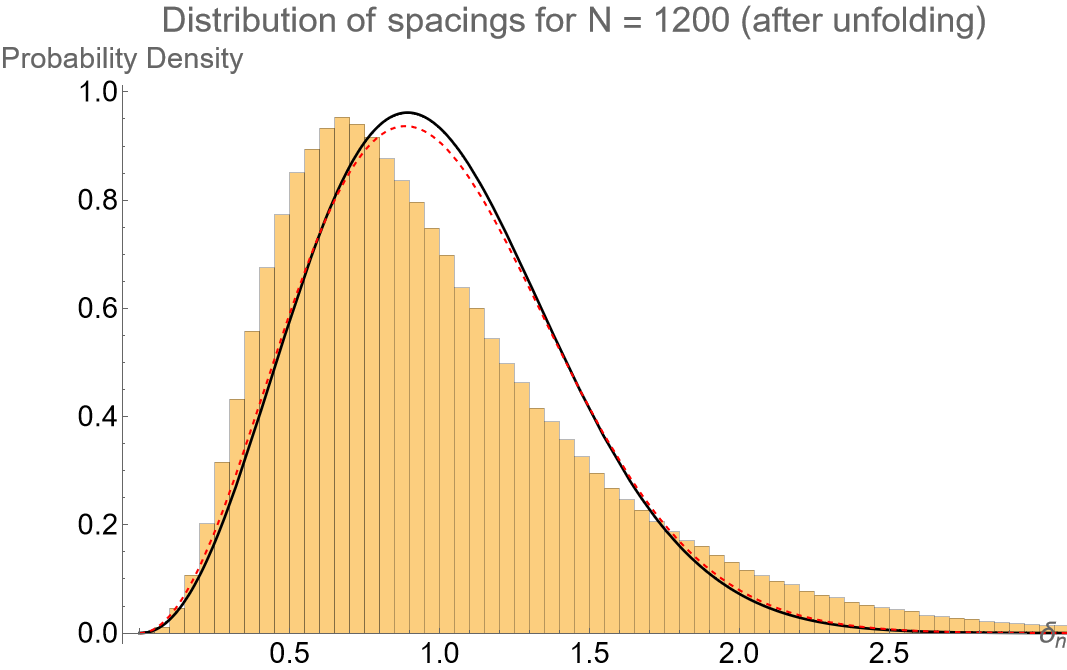}
    \caption{Distribution of spacing ratios $\tilde r_n$ before (top-left) and after unfolding (top-right) for $N=1200$. The fitted distribution (solid black line) goes from $\beta=1.78$ to $2.16$. The red dashed line is the GUE/CUE prediction. On the other hand, the distribution of the spacings $\delta_n$ (bottom), is still not the usual Wigner-Dyson distribution even after the unfolding.}
    \label{fig:unfolded_rn}
\end{figure}

To illustrate the points discussed above, we plot in figure \ref{fig:eigenvaluedensityHTT} the eigenvalue density, in figure \ref{fig:unfolding} an example of the spectrum before and after unfolding, and in figure \ref{fig:unfolded_rn} the distributions of spacings and spacing ratios. In table \ref{tab:unfoldingrn} we list the values of $\avg{\tilde r_n}$ and $\beta$ before and after unfolding, to compare with those found in \cite{Bianchi:2023uby}, and show the improved agreement with CUE (and GUE).

\begin{table}[]
    \centering
    \begin{tabular}{|c|c|c|c|c|} \cline{2-5}
         \multicolumn{1}{c|}{}&  \multicolumn{2}{c|}{Before unfolding} &  \multicolumn{2}{c|}{After unfolding}\\ \hline
         $N$ & $\avg{\tilde r}$ & $\beta$ & $\avg{\tilde r}$ & $\beta$ \\ \hline
         400 &  0.599    &  1.98    &   0.620    & 2.36 \\ \hline
         800 &   0.591  &  1.82    &   0.612    & 2.19 \\ \hline
         1200 &   0.590   & 1.78     & 0.611       & 2.16 \\ \hline
         1600 &  0.588    &  1.74     &  0.610     & 2.12 \\ \hline
    \end{tabular}
    \caption{Results from the distribution for spacing ratios before and after unfolding. The spacing ratios are sensitive to the unfolding procedure, which brings the distribution at large $N$ close to the expected CUE ($\avg{\tilde r} = 0.603$, $\beta = 2$). See figure \ref{fig:unfolded_rn} for plots of the distribution.}
    \label{tab:unfoldingrn}
\end{table}

%\clearpage

\subsection{ScFF for the string three point amplitude}
After performing the unfolding of the ``spectrum'' of $z_n$, the locations of the peaks of the HTT amplitude, we can compute the ScFF.

We will do it separately for ``intermediate'' $N \sim 100$--$1000$, and very large $N \sim 10000$, since the analysis requires different strategies for in the two cases.

\subsubsection{Ensemble averaging for intermediate \texorpdfstring{$N$}{N}}

Since the unfolding procedure and the ScFF are sensitive to the size of the matrices (number of eigenvalues), we pick many states of the string at a given mass level $N$ with fixed number of eigenvalues $L$. This number depends in a highly non-trivial way on the partition of $N$, but it is easy to compute for any given partition. It has a distribution around an average value of $\avg{L} \approx 0.44 N$.

To prepare a sample, we draw many partitions of $N$ at random, and select only those that have a fixed number of eigenvalues $2L$. We pick this number to be near the average/most likely value for a random partition, $2L=\frac{9}{20}N$. Then, we calculate the locations of $z_n$ for each state in the sample. Since there is a symmetry of the amplitude taking $z\to1-z$ ($\alpha\to\pi-\alpha$), we take only the first half of the spectrum with $0 < z_n < \frac12$, and the number of eigenvalues we use is what we denote $L$.

We use the measured \emph{aggregate eigenvalue density} to unfold the obtained spectra as described in the previous section. Then we can compute the ScFF for each state and perform the ensemble averaging by summing over the many states in our random sample.

\begin{figure}[ht!] \centering
     \includegraphics[width=0.48\textwidth]{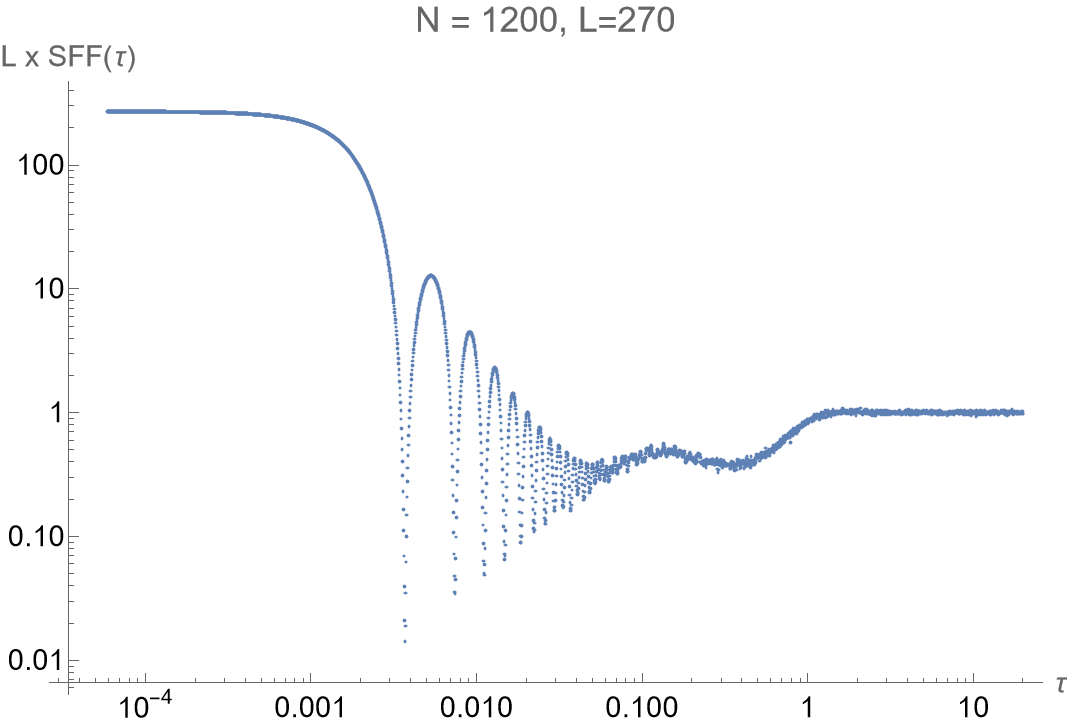}
    \includegraphics[width=0.48\textwidth]{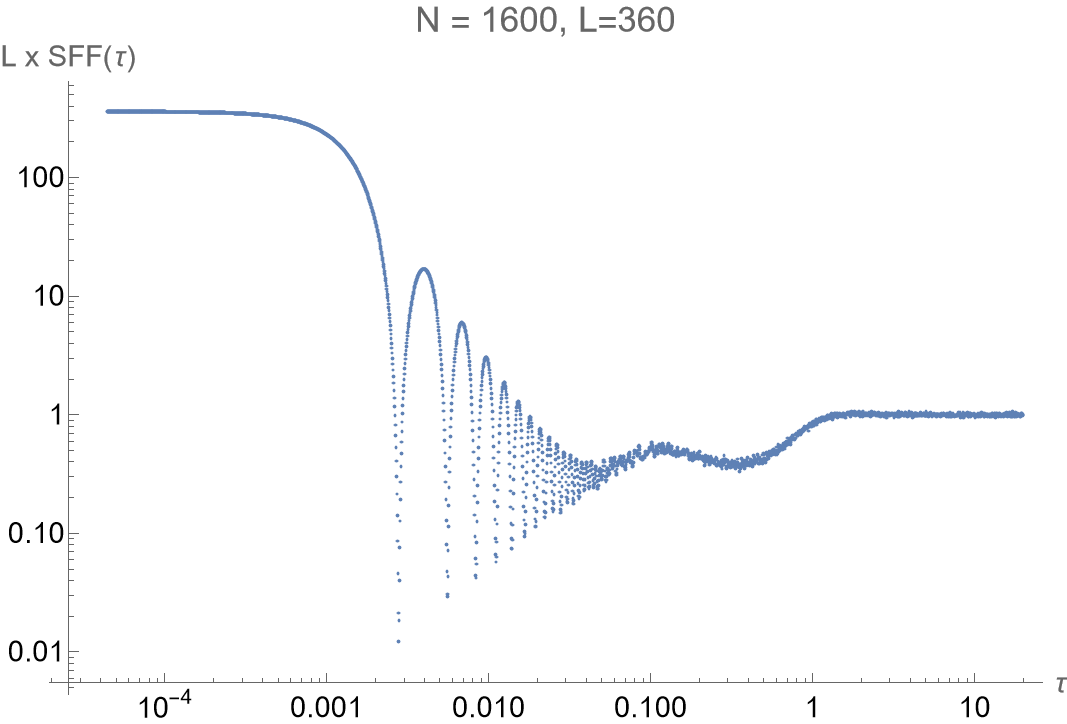}
    \caption{ScFF for the HTT amplitude for unfolded eigenvalues on a logarithmic scale, showing four distinct regions: decline, bump, ramp and plateau.}
    \label{fig:sff_string_loglog}
\end{figure}

\begin{figure}[ht!] \centering
    \includegraphics[width=0.48\textwidth]{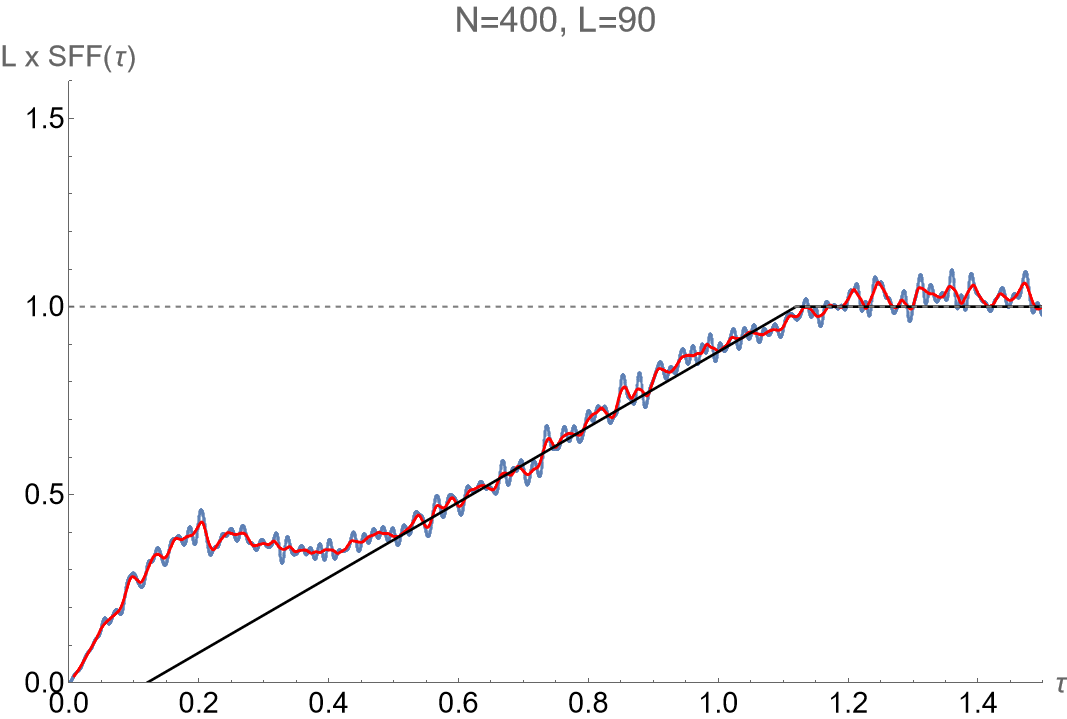}
    \includegraphics[width=0.48\textwidth]{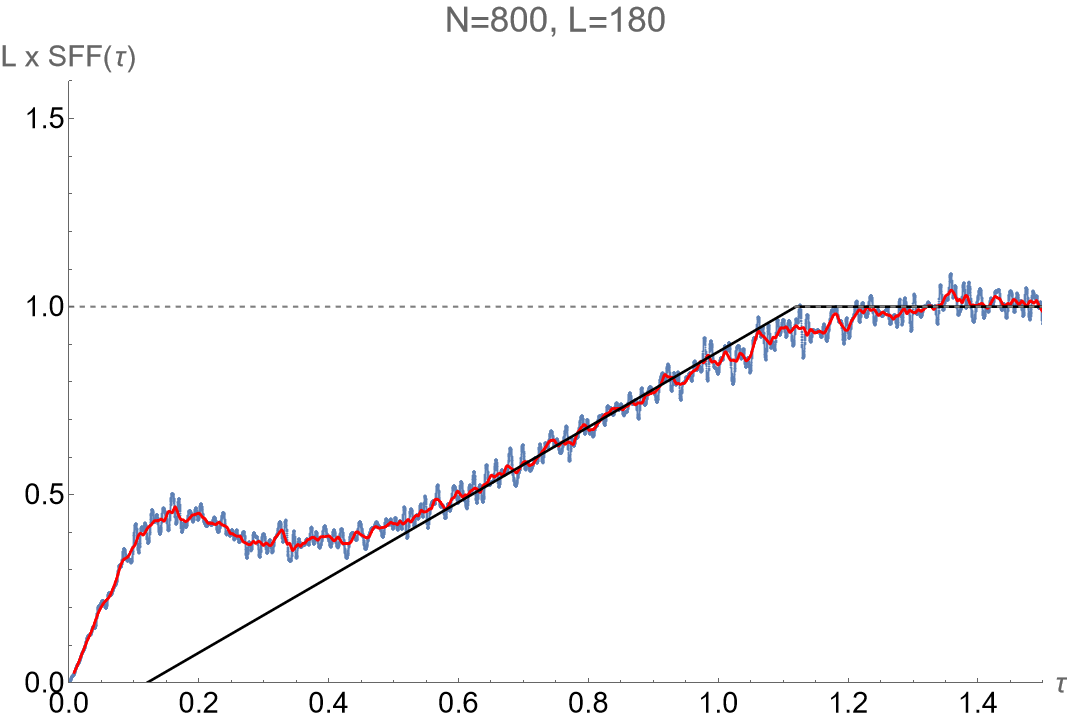} \\
    \includegraphics[width=0.48\textwidth]{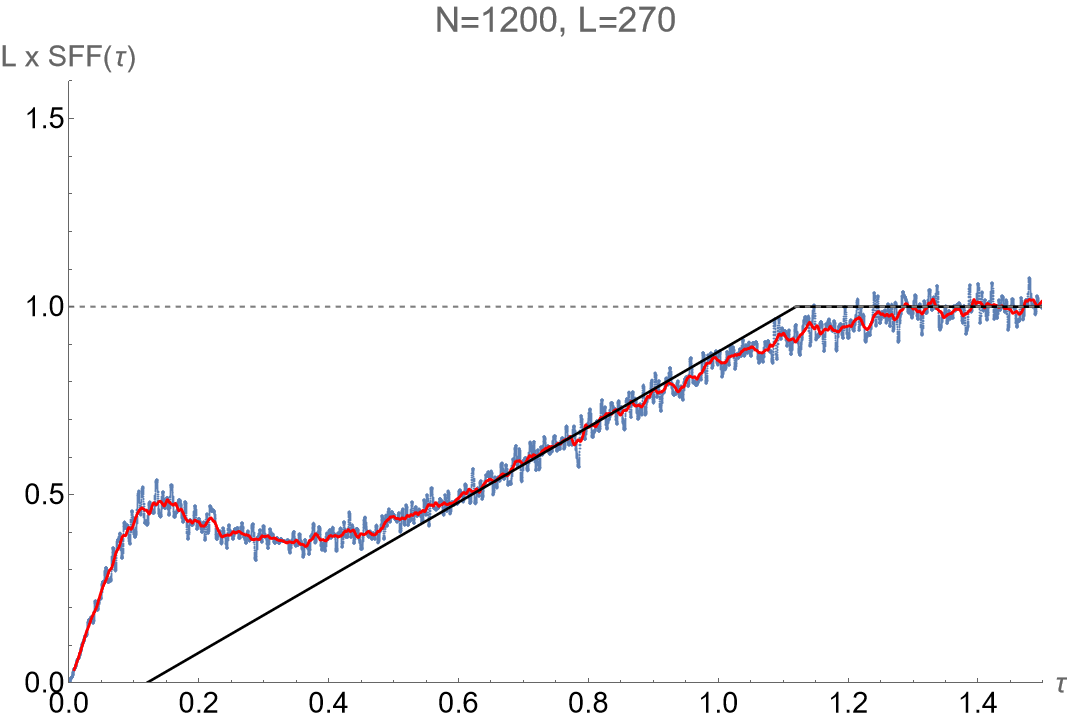} 
    \includegraphics[width=0.48\textwidth]{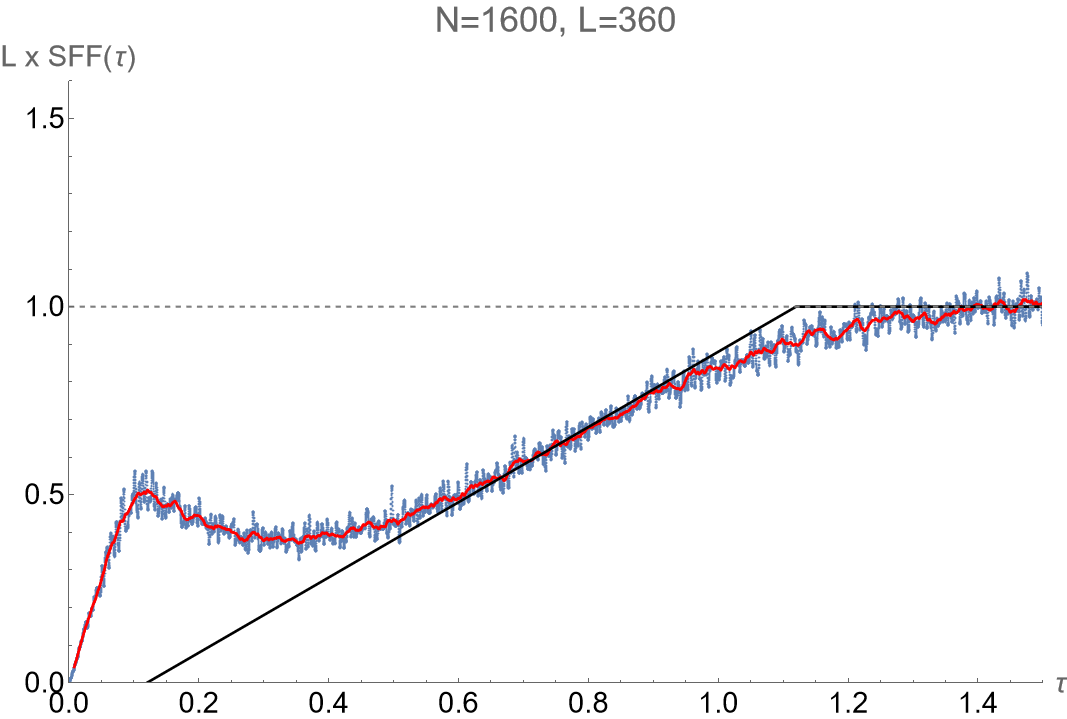}
    \caption{Connected part of the ScFF for the HTT amplitude for unfolded eigenvalues, computed over a sample of 2000 states with fixed $L$ at each level $N$. The blue points are the ScFF, the red line is their time average. The result is compared with a (delayed) linear ramp with the CUE slope (black line).}
    \label{fig:sff_string_connected}
\end{figure}

The resulting ScFFs are plotted in figures \ref{fig:sff_string_loglog} (logarithmic scale) and \ref{fig:sff_string_connected} (connected part, linear scale). We see the classic universal features of chaotic systems: decline to a dip, a linear ramp, and eventually a plateau, consistent with RMT, with some modifications.

There is a marked linear ramp with a slope consistent with CUE, but at early times there is a noticeable ``bump'' before the ramp. Figure \ref{fig:sff_string_loglog} shows that there in fact four distinct regions: dip, bump, ramp, plateau, each happening at a different time.

Furthermore, the ramp begins at a later time and appears to be shifted by a constant. Using the same notation as in section \ref{sec:rmtsff}, the ramp appears to have the form
\be r_2(\tau)\approx \tau-\Delta\tau \ee
instead of being simply $r_2(\tau) = \tau$. There is a delay time shifting the ramp and the plateau by $\Delta\tau\approx 0.12$--0.15.

It is difficult to measure precisely the slope of the ramp from the results of figure \ref{fig:sff_string_connected}. As noted in section \ref{sec:rmtsff}, we expect at small times a linear ramp in the connected part of the ScFF that goes like $2\tau/\beta$, but here this region is dominated by the bump structure. At later times, the ramp is linear (consistent with CUE), or slightly concave ($\beta<2$). The best overall fit seems to be taking the modified linear formula with the shift $\Delta\tau$. The $N$ dependence is weak in the range examined (400--1600).

In the following subsection, we show that the bump can be eliminated when considering much larger $N$, when we have much larger sets of eigenvalues per state, and can account for the specific structure of the state considered when unfolding.

% \begin{figure}[ht!] \centering
%     \includegraphics[width=0.48\textwidth]{figSFF/sff_string_400.png}
%     \includegraphics[width=0.48\textwidth]{figSFF/sff_string_800.png} \\
%     \includegraphics[width=0.48\textwidth]{figSFF/sff_string_1200.png}
%     \includegraphics[width=0.48\textwidth]{figSFF/sff_string_1600.png}
%     \caption{ScFF for the HTT amplitude for unfolded eigenvalues, computed over a sample of 2000 states with fixed $L$ at each level $N$. The blue points are the ScFF, the red line is their time average. The result is compared with a (delayed) linear ramp with the CUE slope (black line).}
%     \label{fig:sff_string}
% \end{figure}

\subsubsection{ScFF and spacing ratio analysis at very large \texorpdfstring{$N$}{N}}
For very large $N\sim10000$, we have enough data from any single amplitude of a generic HES state with two tachyons to compute the distribution of spacings of peaks and the corresponding ScFF. We can see that they display all the features of chaotic systems.

Recall that the number of poles in the amplitude is equal to the number of eigenvalues plus one, and that the placement of these poles is dependent on the integers $n$ that appear in the partition defining the string state. While the number of poles is always linear in $N$, their placements can be very different for different partitions of $N$, and with large $N$ many different values of $n$ can appear in different partitions.  As argued above this introduces a sensitivity of the results to the precise unfolding process.

We have collected results for the spectra of $N = 10000$ (3 states), 20000 (3 states), and 40000 (a single state). The values of the helicity are fixed to the most likely value at $h\approx\frac{\sqrt6}{2\pi}\sqrt{N}\log N$. We do not fix the number of eigenvalues $L$, so this varies slightly for each state and is around $2L\approx N/2$ for the states considered.

\begin{figure}
    \centering
    \includegraphics[width=0.60\textwidth]{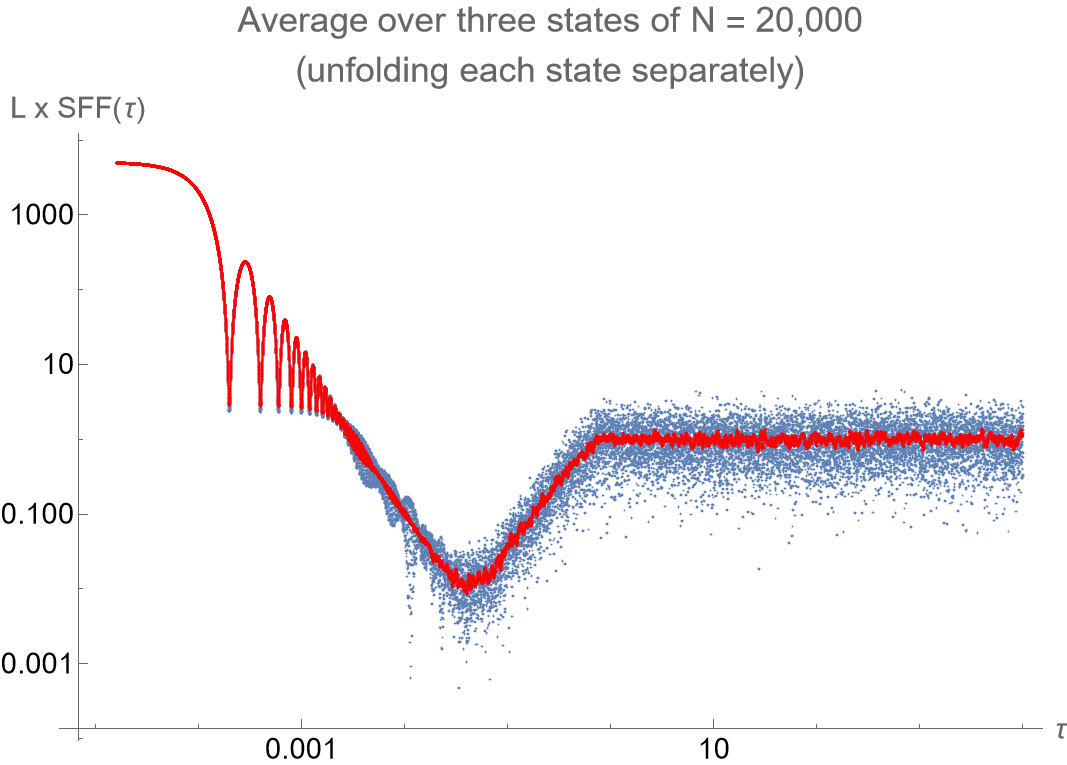} \\
    \includegraphics[width=0.40\textwidth]{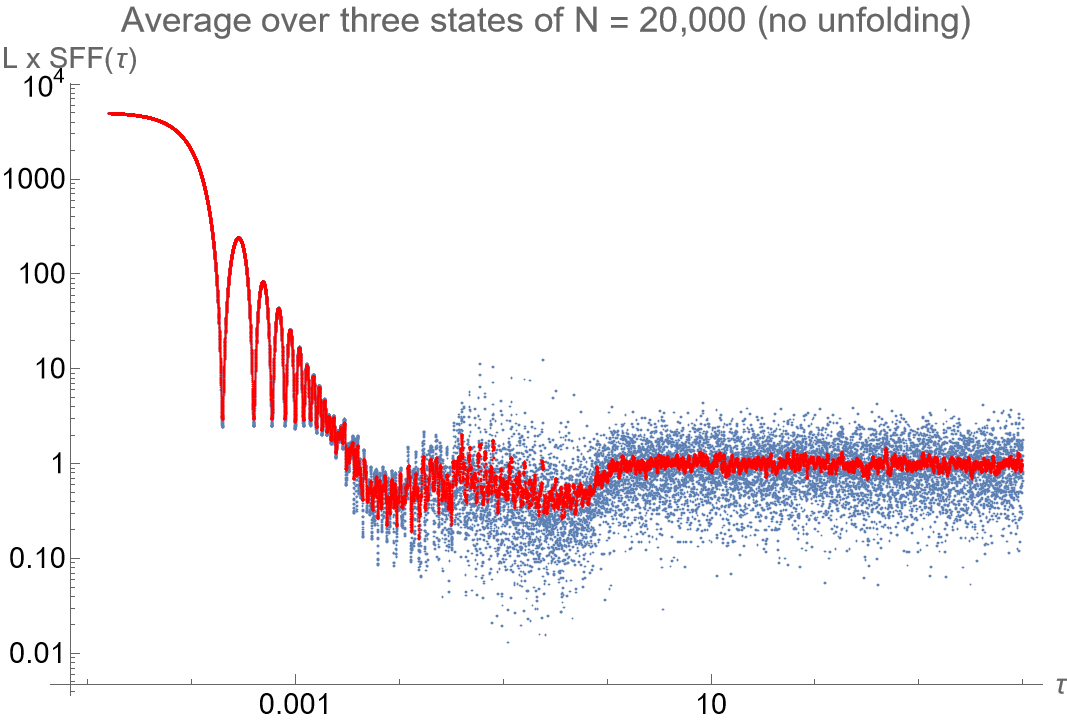}
    \includegraphics[width=0.40\textwidth]{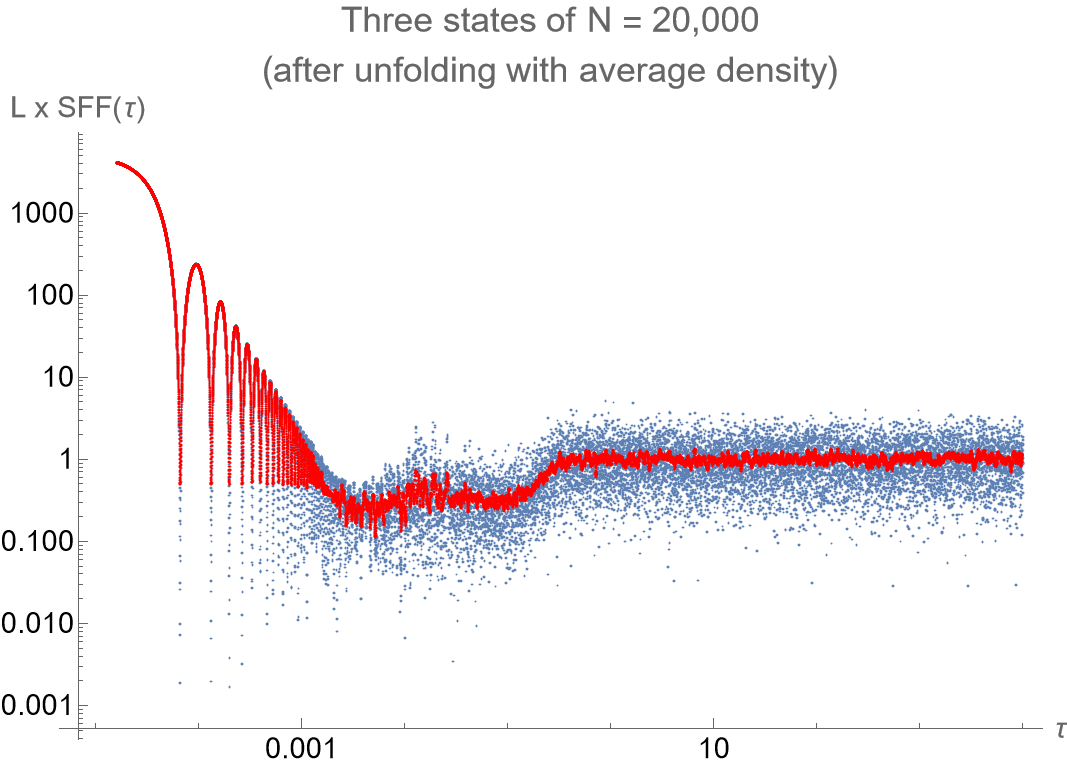}
    \caption{Results for the ScFF on a logarithmic scale computed from three states of $N=20,000$ and $h=519$. In the top figure, each state had its spectrum unfolded using its own eigenvalue density function, and then the three ScFF were averaged. In the bottom figures are displayed for comparison the results for the spectra without unfolding (left), and the spectra when unfolded with a single average density function computed by aggregating the results of the three chosen states (right). These display a ``bump'' before the ramp.}
    \label{fig:sff_string_largeN_loglog}
\end{figure}

For each given state, we compute the positions of the peaks of the amplitude, from it we obtain the eigenvalue density empirically and unfold the spectrum. For the resulting spectra we check the distributions of spacings and spacing ratios and compute the ScFF. We see that when the spectra are unfolded for each state with its own density function, the ``bump'' in the ScFF disappears, and we see the classic dip-ramp-plateau structure. See figure \ref{fig:sff_string_largeN_loglog}.

\begin{figure}
    \includegraphics[width=0.48\textwidth]{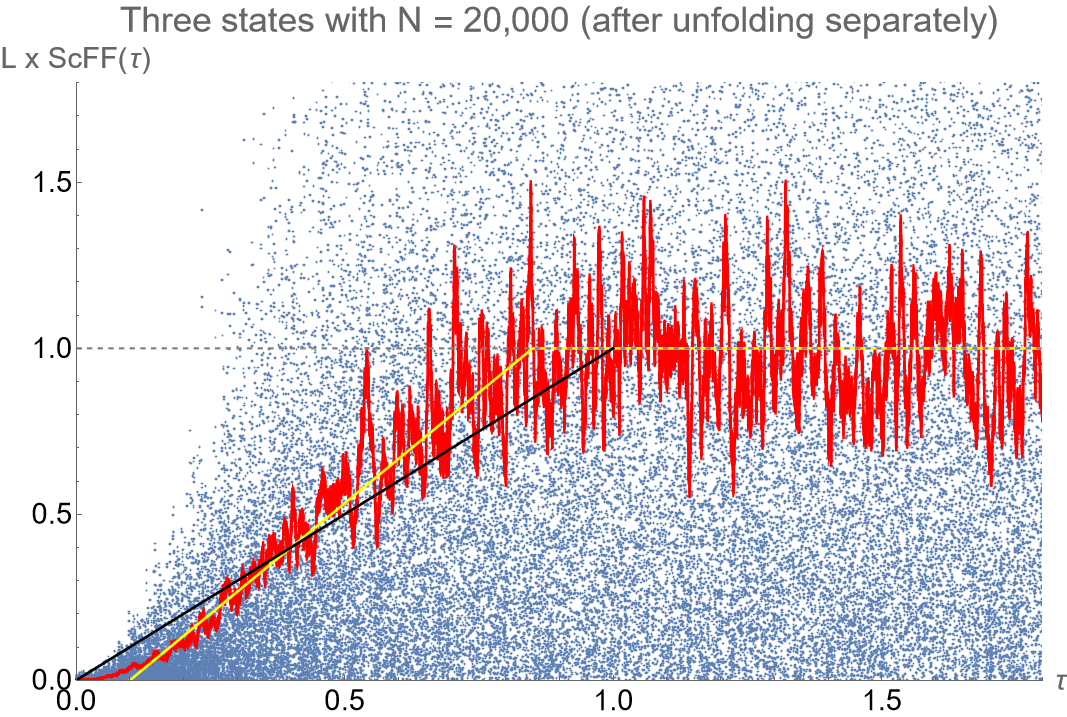}
    \includegraphics[width=0.48\textwidth]{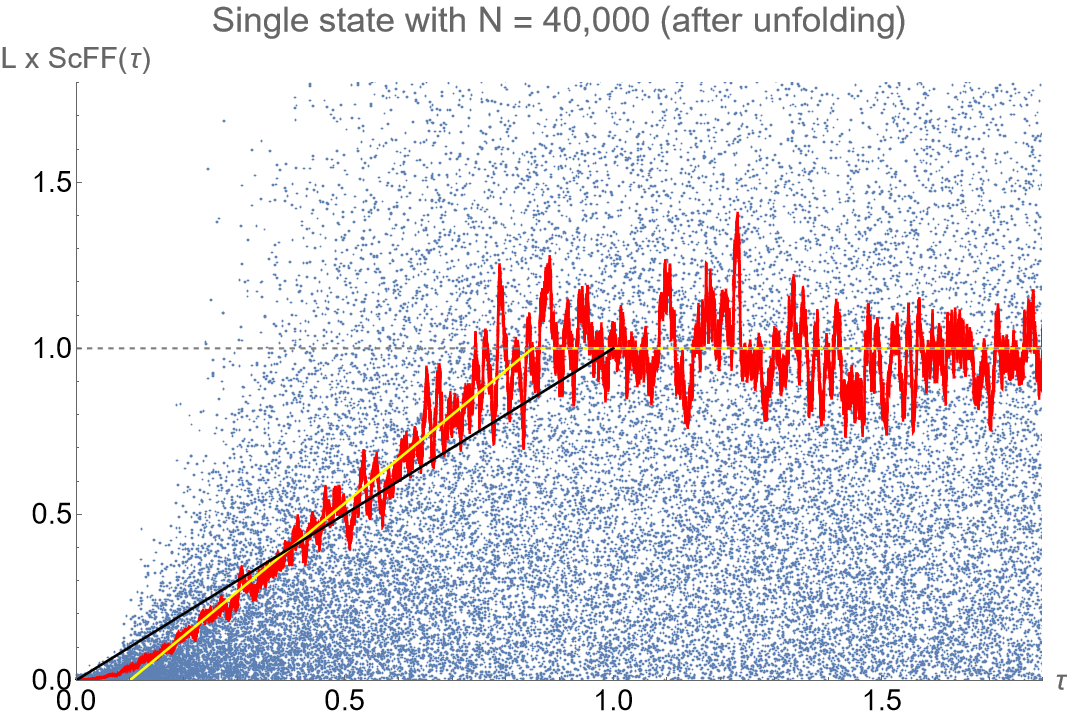}
    \caption{Connected part of the ScFF for three states of $N=20000$ (each unfolded separately), and one state of $N=40000$. We compare with the CUE prediction (solid black line), and a linear ramp with a time delay and shift (yellow line).}
    \label{fig:sff_string_largeN_connected}
\end{figure}

There is still a time delay before the linear ramp. See figure \ref{fig:sff_string_largeN_connected}. The ramp fits best the linear formula
\be r_2(\tau) = \frac{\tau-\Delta\tau}{\tau_p-\Delta\tau} \ee
The apparent linearity of the ramp suggests the best agreement is with CUE, though we need to introduce by hand the delay time $\Delta\tau \approx 0.1$ and shift the plateau to $\tau=\tau_p\approx 0.9$, which rescales the $\tau$ variable and the slope of the ramp. In any case, a linear ramp fits the results better than any $\beta$-ensemble with value different from 2.

\begin{figure}
    \includegraphics[width=0.48\textwidth]{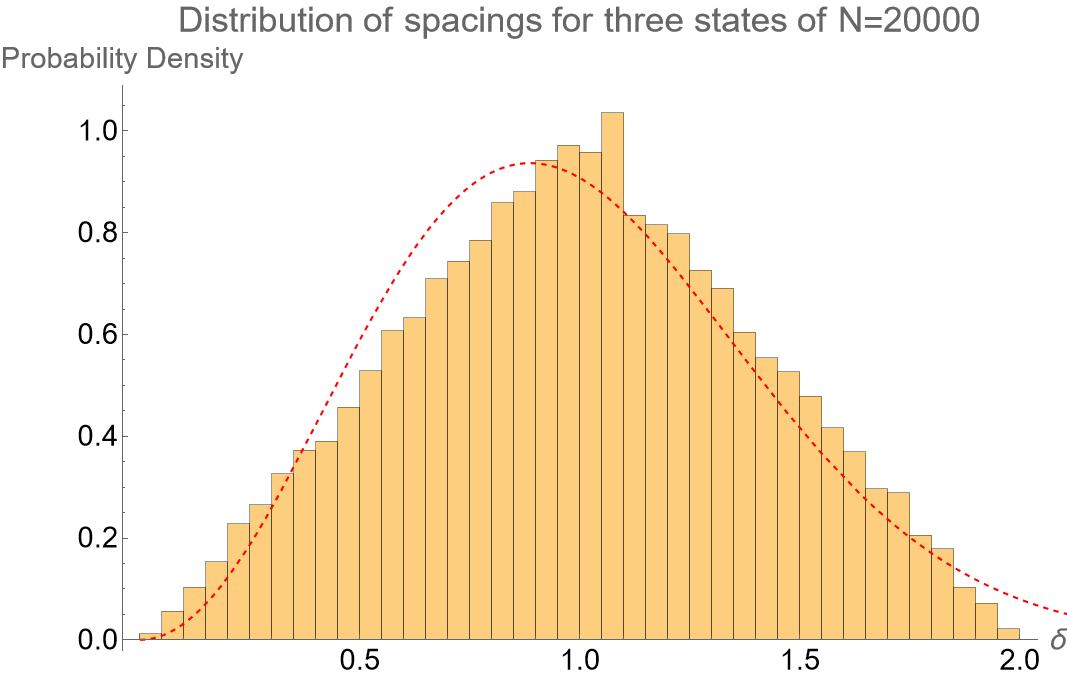}
    \includegraphics[width=0.48\textwidth]{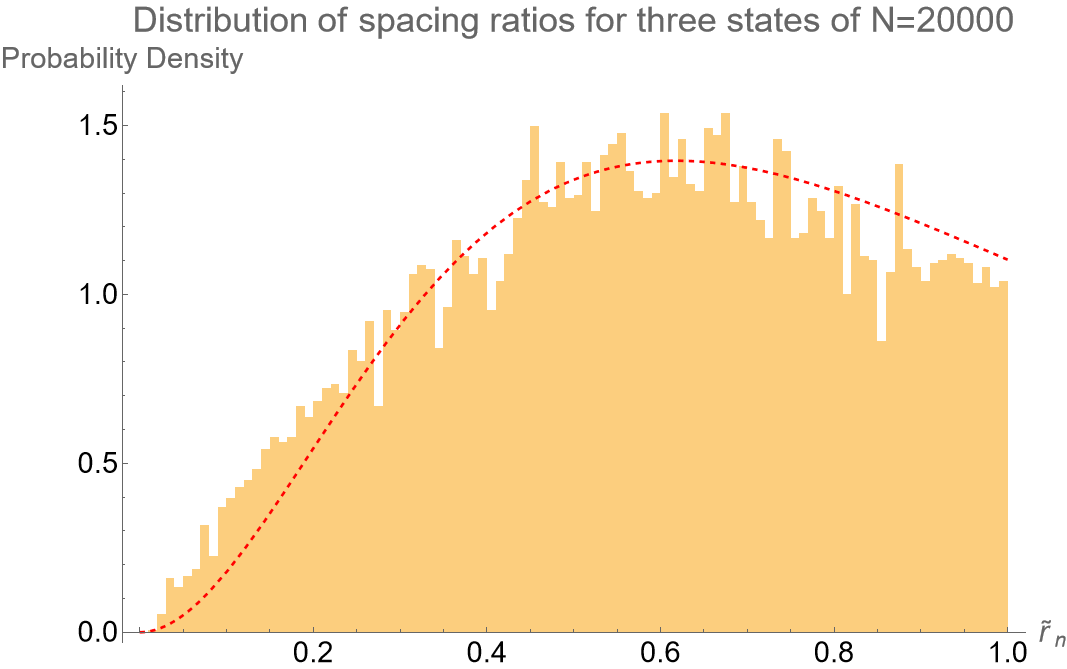}
    \caption{Distributions for the spacings $\delta_n$ and the ratios $\tilde r_n$ for three states of $N=20000$ (each unfolded separately before combining the results). The results agree with the CUE prediction (dashed red line).}
    \label{fig:largeN_spacings}
\end{figure}

Finally, the distributions of spacings and spacing ratios for the three states at $N=20000$ can be seen in figure \ref{fig:largeN_spacings}. We plot the results against the CUE formulas (Wigner-Dyson distribution). There are deviations, but the spacings show better agreement than for the lower $N$ states examined before. We expect the agreement to improve when taking more states at this level and subjecting them to the same analysis.

%\clearpage

\section{Summary}
We have introduced the concept of a ``scattering form factor'' (ScFF), dubbed in consonance with the celebrated spectral form factor (SFF), and proposed it as a novel useful indicator for the chaotic behavior of scattering amplitudes, to be used in conjunction with the distribution of spacings of peaks as proposed in \cite{Bianchi:2022mhs,Bianchi:2023uby}.

We have spelled out the main properties of the ScFF, and illustrated them in several instances of integrable models, and the two examples of chaotic scattering: the Gutzwiller leaky torus and the scattering amplitudes of highly excited string (HES) states. 

We have refined our analysis of the chaotic behaviour of two-body decays of HES states at tree level for open bosonic strings. In particular, by unfolding the spectrum of eigenvalues, which are the zeros of the logarithmic derivative of the amplitude, we have found a behaviour very close to the one of GUE/CUE with a best fit of $\beta$ around 2. On top of this we have found a ``bump'' after the decline that we interpreted as due to the presence of repulsive points in the angular spectrum, where ${\cal F}$ has poles and cannot have zeros. However we have shown that this bump structure disappears after the correct unfolding of the spectrum.

Let us elaborate further on these repulsive points. They owe their presence to the extreme regularity of the free string spectrum and possibly to its large degeneracy at fixed level. Turning on interactions we expect the states at level $N$ to spread around and acquire imaginary parts in their masses to be associated to their width, i.e. their finite life-time, so that the repulsive points would be less or even non-repulsive. This is reminiscent of ``scar states'' that correspond to integrable trajectories present in some chaotic models \cite{Dodelson:2022eiz, Dodelson:2023vrw}. These seem to violate the Eigenstate Thermalization Hypotesis \cite{Srednicki_1994} and thus give rise to `echos' or revivals in observables that can be approximated by the SFF or our ScFF depending on the context. We plan to come back to this point in the future.

We end on a few comments and observations. First, a similar analysis can be performed for different observables other than the angular distribution. In fact the distribution of the eigen-phases of the $S$-matrix for Compton-like scattering of low-mass probes off HES states for open and closed bosonic strings at tree level has been considered recently in \cite{Savic:2024ock}. 

Second, the additional features, {\it e. g.} the `bump', that we found (in string decays) and dispose of by unfolding the spectrum may disappear after including $g_s$ corrections to the spectrum that would shift the poles and remove all or part of the exponential degeneracy $d_N \sim e^{C\sqrt{N}}$ at large $N$. A quantitative analysis of this effect is well beyond of the present investigation but it is tempting to speculate that RMT would play a significant role in this endeavour, very much as in the spectrum of nuclear resonances as shown by Wigner. 

There are plenty of open questions related to the novel proposal of the scattering form factor and its applications. Here we list few of them:
\begin{itemize}
\item 
The map of certain characteristic properties  of chaotic spectra to those  associated with decays and scattering processes raises a broader question regarding the mapping of general $S$-matrix properties to those associated with Hamiltonians of physical systems. In particular one should pursue a more rigorous understanding of the ScFF variable $s$ that replaces the time in the SFF and that we argued to be associated to (complexified) angular momentum or partial wave (when the angular dependence of an amplitude is considered).
\item 
The set of characteristic features of RMT are not only the eigenvalues of the random matrices but also the corresponding eigenvectors. Both the ratios of the spacings and the form factors are based only on the former and not on the latter. Relating the eigenvectors of the random matrices and the values of the scattering amplitudes at their maxima points is an obvious task for the correspondence between chaotic scattering amplitudes and RMT, and is important in the context of the famous eigenstate thermalization hypothesis. The spectra of certain (chaotic) QFTs were examined in this view in \cite{Srdinsek:2020bpq} and \cite{Delacretaz:2022ojg} and were found to be problematic.

\item 
In \cite{Bianchi:2022mhs, Bianchi:2023uby} and in the present paper we have proposed quantitative measures  for the chaotic behavior of HES decays and scattering processes, however the mechanism that yields this type of behavior is still to be deciphered. In particular the transition from non-chaotic behavior for small values of $N$ and $h$ (or for $h\sim N$) to a chaotic one for large values of $N$ and generic $h$, requires further study. 

\item 
We have shown that the ScFF of the phase shifts associated with the leaky torus is in one to one correspondence with the SFF (or function-zero FF) built from the zeros of the Riemann zeta function. The case of the leaky torus can be generalized to other topologies and geometries and one could speculate that these could correspond to generalizations of the Riemann zeta function.

\item
We have observed a ``bump" in the ScFF that generically does not exist in the spectral case since it can be disposed of by unfolding the spectrum. We also conjectured that this novel phenomenon follows from the voids in the spectrum of the peaks of the scattering amplitudes, which exists because at certain angles the amplitude is zero and cannot have a peak.  It will be interesting to look for bumps or similar deviations from standard chaotic behaviour in other scattering processes, in the form factors associated with spectra of other Hamiltonian systems, to zeros of other mathematical functions, or to find their analogies in RMT ensembles other than the standard ones.
\end{itemize}

%Papers on strings and chaos

\section*{Acknowledgments}
We thank F. Fucito, J. F. Morales, A. Gaikwad, N. Shrayer, C. Sleight, M. Taronna, J.G. Yoon, C.A. Rosen, V. Niarchos and E. Kiritsis for useful discussions.

MB would like to thank the MIUR PRIN contract
2020KR4KN2 ``String Theory as a bridge between Gauge Theories and Quantum Gravity" and
the INFN project ST\&FI ``String Theory and Fundamental Interactions" for partial support.

The work of MF is partially supported by the European MSCA grant HORIZON-MSCA-2022-PF-01-01 ”BlackHoleChaos” No.101105116 and by the H.F.R.I call “Basic research Financing (Horizontal support of all Sciences)” under the National Recovery and Resilience Plan “Greece 2.0” funded by the European Union – NextGenerationEU (H.F.R.I. Project Number: 15384.).
 MF would like to thank the Nordic Institute for Theoretical Physics (Nordita) for their hospitality during a period when part of the research work has been conducted.

The work of JS was supported in part by a grant 01034816 titled “String theory reloaded- from fundamental questions to applications” of the “Planning and budgeting committee”.

DW was supported by an appointment to the YST Program at the APCTP through the Science and Technology Promotion Fund and Lottery Fund of the Korean Government. This was also supported by the Korean Local Governments of Gyeongsangbuk-do Province and Pohang City.

%%\clearpage

\appendix

\section{Integer partitions and structure of peaks in the HES decay amplitude} \label{sec:numzeros}
To restate the problem we want to address here (introduced in the main text in section \ref{sec:unfolding}), the number of peaks of the amplitude of one HES with two tachyons, or number of eigenvalues of the corresponding random scattering matrix, depends on the state discussed, i.e. on the specific partition of the level $N$.

It is equal to one plus the total number of unique fractions of the form $\frac{k}{n}$, where $n$ is one of the numbers that appear in the partition, and $k = 1,2,\ldots n-1$.

A partition is represented as a list $\{g_n\}$, $n = 1,2,\ldots N$, where $g_n$ is the number of times that $n$ occurs in the partition. There is a well-known result \cite{Fristedt:1993} for the asymptotic distributions of $\{g_n\}$ for large $N$, namely that each $g_n$ has the geometric distribution
\be P(g_n = k) = (1-p_n)^k p_n \label{eq:distnm} \ee
with the parameter
\be p_n = 1 - \exp\left(-\frac{ n \pi}{\sqrt{6 N}}\right) \ee
This is correct on the sets of $\{g_n\}$ satisfying the basic constraint that each set represents a partition of $N$: $\sum_{n} n g_n = N$.\footnote{This distribution is also the basis of the probabilistic algorithm that we used to generate random partitions of large $N$ (see \cite{Arratia:2016} for the original reference, and appendix D of our previous paper \cite{Bianchi:2023uby} for the details of implementation.}

The parameter $p_n$ is also equal to the probability that the number $n$ appear at least once in a partition. Therefore the expectation value of $g_n$ in a partition of $N$ is
\be \langle g_n \rangle_N = \frac{1-p_n}{p_n} = \frac{1}{e^{n \pi /\sqrt{6 N}}-1} \ee
which we immediately recognize as a Bose-Einstein distribution with energy levels $\varepsilon_n = n\pi$ and temperature $T = \sqrt{6N}$.

Now, the probability that a certain pole $z_i$ appear in ${\cal F}$ is as follows. If $z_i$ is given in its irreducible form $z_i=k/q$, and $m_q = \lfloor N/q \rfloor $,
%\be \langle r_i \rangle = \sum_{n/q} \langle g_n \rangle = \sum_{n/q} \frac{1-p_n}{p_n} \ee
the probability that $z_i = k/q$ will appear is the probability that any multiple of $q$ will appear at least once in the partition. It is equal to
\be \tilde p_q = 1 - \prod_{k=1}^{m_q}\left(1-e^{-\frac{k q \pi}{\sqrt{6N}}}\right)\ee

For a given $n$ the number of fractions of the form $z = k/n$, $k=1,\ldots n-1$, that are irreducible exactly matches the definition of $\varphi(n)$, Euler's totient function which counts how many integers smaller than $n$ are co-prime to $n$. 

If a number $n$ appears in a partition, then it will necessarily contribute $\varphi(n)$ new poles to ${\cal F}$, and therefore the expected number of zeroes for a partition of $N$ is 
\be \avg{\text{Number of zeros}} = 1 + \sum_{n=1}^{N}\varphi(n) \tilde p_n \ee
At large $N$ this asymptotes to $\approx 0.44 \times N$. This can also be verified by the experiment of computing the average number of zeros over a random sample of many partitions of $N$. 

\begin{figure}
    \centering
    \includegraphics[width=0.60\textwidth]{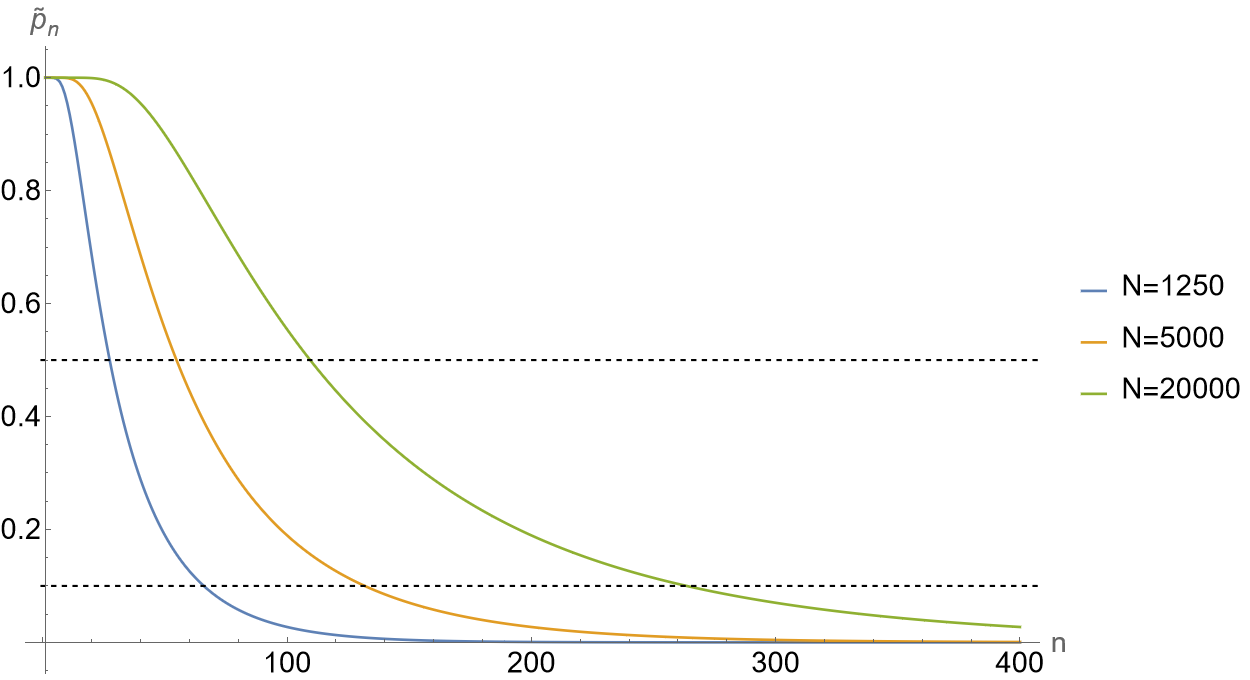} \\
    \includegraphics[width=0.60\textwidth]{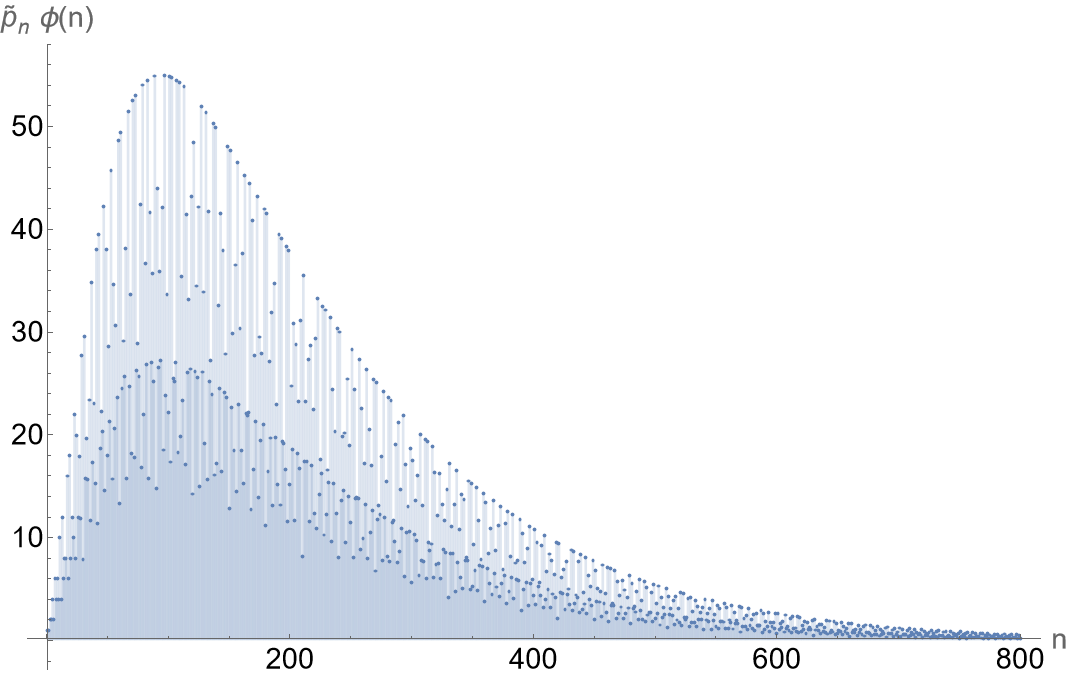}
    \caption{Left: plot of $\tilde p_n$ as a function of $n$ for different values of $N$, showing the probability that $z=k/n$ is a pole in an amplitude of that level. The horizontal lines are drawn at $p = \frac12$ and $\frac1{10}$. Right: plot of $\tilde p_n$ as a function of $n$ for $N=20000$, showing the expected number of poles that each contributes to an amplitude for a random partition of $N$.}
    \label{fig:pn}
\end{figure}

We illustrate the probabilities in figure \ref{fig:pn}.

% We can also write an expression for the strength of the ``repulsion'' of eigenvalues from a certain pole $z_i = k/q$. The residue at $z=z_i$ in the logarithmic derivative of the HTT amplitude, eq. \eqref{...}, is the sum of $g_n$ such that $n$ is a multiple of $q$. It is also the degree of the zero $z_i$ in the amplitude itself. We can compute the expectation value of it as
% \be \sum_{k=1}^{m_q} \avg{g_{k q}} = \sum_{k=1}^{m_q} \frac{1}{e^{\pi k q/\sqrt{6N}}-1} \ee
% This doesn't seem correct, maybe because it fails to account for the constraint on $g_n$??

\section{Low levels spin decomposition} \label{sec:decomposition}
To get an idea of the origin of chaotic behavior in the HES amplitudes, we can decompose the HTT amplitude \eqref{eq:A_HTT} in terms of the associated Legendre polynomials, $P_l^m \equiv P_{l,m}$, to get their spin decomposition.

For a state with given $N$ and $h$, because $h$ is the helicity of the state, the decomposition is in terms of $P_{s,h}(\cos\alpha)$ with the spin $s$ naturally constrained to be $h\leq s \leq N$.

Recall also that we have chosen a specific subset of excited states that can be constructed in the DDF approach using only photons with the same circular polarization $\lambda^2=0$, and that the angle of an outgoing tachyon $\alpha$ is defined relatively to the direction of the DDF photon's momentum.

For small $N$, we can do it for all partitions easily. There is a common factor to all amplitudes, so lets define for each partition
\be {\tilde A} = \frac{(-1)^N(2N-1)!!}{\pi^h} A \ee
with the double factorial $(2N-1)!! = 1\times3\times5\times7\times\ldots\times(2N-1)$.

Representing a partition of $N = n_1+n_2+\ldots + n_h$ as the list of $\{n_i\}$, we write the decomposition
\be \tilde A = \sum_{s=h}^N c_{s} P_{s,h} \ee
below for each partition, for the first few levels:
\begin{itemize}
    \item For $N=1$,
        \be \{1\} =  P_{1,1} \nonumber \ee
    \item For $N=2$,
        \begin{align*}
        \{1,1\} &= P_{2,2} \\
        \{2\} &=  P_{2,1}
        \end{align*}
    \item For $N=3$,
        \begin{align*}
        \{1,1,1\} &= P_{3,3} \\
        \{2,1\} &=  P_{3,2} \\
        \{3\} &=  \frac32 P_{1,1} + \frac94 P_{3,1} \\
        \end{align*}
    \item For $N=4$,
        \begin{align*}
        \{1,1,1,1\} &= P_{4,4} \\
        \{2,1,1\} &=  P_{4,3} \\
        \{2,2\} &=  5 P_{2,2} + 2 P_{4,2} \\
        \{3,1\} &=  \frac54 P_{2,2} + \frac94 P_{4,2} \\
        \{4\} &=  \frac{25}{3} P_{2,1} + 8 P_{4,1}
        \end{align*}
    \item For $N=5$,
        \begin{align*}
        \{1,1,1,1,1\} &= P_{5,5} \\
        \{2,1,1,1\} &=  P_{5,4} \\
        \{2,2,1\} &=  7 P_{3,3} + 2 P_{5,3} \\
        \{3,1,1\} &=  \frac94 P_{5,3} \\
        \{3,2\} &=  \frac{63}{4} P_{3,2}+\frac{27}{4} P_{5,2} \\
        \{4,1\} &=  7 P_{3,2}+8 P_{5,2} \\
        \{5\} &=  \frac{495}{16}P_{1,1} + \frac{875}{16}P_{3,1} + \frac{625}{16}P_{5,1}
        \end{align*}
\end{itemize}
The amplitude of the partition with $h=N$ is always $P_{N,N}$, and likewise the single state with $h=N-1$ is just $P_{N,N-1}$. For the partition $\{N\}$ with $h=1$, the last term is always
\[ \{N\} = \ldots + \left(\frac{N}{2}\right)^{N-1} P_{N,1} \]

There is also a constraint, such that only even (odd) spins appear for even (odd) $N$. All coefficients turn out to be positive for the lowest levels, but going to larger $N$ we see that for some partitions, we get negative coefficients as well, only for some of the partitions. 

We have attempted to study the distribution of the coefficients in this decomposition for higher $N$, but it remains an open question how one can extract additional data from it on the origin and nature of chaotic behavior in the HTT amplitude.

\bibliographystyle{JHEP}
\bibliography{main}

\end{document}